\documentclass[journal]{IEEEtran}
\usepackage{graphicx}
\usepackage{subfigure}
\usepackage{float}
\usepackage{mathrsfs}
\usepackage{amsfonts}
\usepackage{array}
\usepackage{amssymb,amsmath}
\usepackage{longtable}
\usepackage{rotating}
\usepackage{multirow}
\usepackage{epstopdf}
\usepackage[lined,boxed,commentsnumbered, ruled]{algorithm2e}
\usepackage{cite}
\usepackage{float}
\usepackage{stfloats}
\usepackage{algorithmic}
\usepackage[colorlinks,
            linkcolor=black,
            anchorcolor=black,
            urlcolor=blue,
            citecolor=black]{hyperref}
\usepackage[hyphenbreaks]{breakurl}

\renewcommand{\baselinestretch}{1}

\def\BibTeX{{\rm B\kern-.05em{\sc i\kern-.025em b}\kern-.08em
    T\kern-.1667em\lower.7ex\hbox{E}\kern-.125emX}}

\renewcommand{\baselinestretch}{1}
\usepackage{color}
\usepackage{enumitem}
\usepackage{booktabs}

\begin{document}

\title{ACSNet: A Deep Neural Network for Compound GNSS Jamming Signal Classification}
\author{Min Jiang,\IEEEmembership{~}
Ziqiang Ye,\IEEEmembership{~}
Yue Xiao,\IEEEmembership{~Member, IEEE,}
Yulan Gao,\IEEEmembership{~Member, IEEE,}
Ming Xiao,\IEEEmembership{~Senior Member, IEEE,}\\
and Dusit Niyato,\IEEEmembership{~Fellow, IEEE}
\thanks{M. Jiang, Z. Ye and Y. Xiao are with the National Key Laboratory of Wireless Communications, University of Electronic Science and Technology of China (UESTC), Chengdu 611731, China (e-mail: minjiang060@gmail.com; yysxiaoyu@hotmail.com; xiaoyue@uestc.edu.cn).}
\thanks{Y. Gao and M. Xiao are with the Division of Information Science and Engineering, KTH Royal Institute of Technology, 100 44 Stockholm, Sweden (e-mail: yulang@kth.se; mingx@kth.se).}
\thanks{Dusit Niyato is with the College of Computing and Data Science, Nanyang Technological University, Singapore (e-mail: dniyato@ntu.edu.sg).}}
\maketitle

\begin{abstract} 
In the global navigation satellite system (GNSS), identifying not only single but also compound jamming signals is crucial for ensuring reliable navigation and positioning, particularly in future wireless communication scenarios such as the space-air-ground integrated network (SAGIN). However, conventional techniques often struggle with low recognition accuracy and high computational complexity, especially under low jamming-to-noise ratio (JNR) conditions. To overcome the challenge of accurately identifying compound jamming signals embedded within GNSS signals, we propose ACSNet, a novel convolutional neural network designed specifically for this purpose. Unlike traditional methods that tend to exhibit lower accuracy and higher computational demands, particularly in low JNR environments, ACSNet addresses these issues by integrating asymmetric convolution blocks, which enhance its sensitivity to subtle signal variations. Simulations demonstrate that ACSNet significantly improves accuracy in low JNR regions and shows robust resilience to power ratio (PR) variations, confirming its effectiveness and efficiency for practical GNSS interference management applications.
\end{abstract}
\begin{IEEEkeywords}
Global navigation satellite system (GNSS), compound jamming signal, convolutional neural network, low JNR, PR variation.
\end{IEEEkeywords}

\section{Introduction}
\IEEEPARstart{W}{ith} the development of the next-generation communication system, the unique advantages of the space-air-ground integrated network (SAGIN) are exhibited as integrating space, air, and ground networks into a unified architecture \cite{xiao2024space}, towards more flexible, efficient, and intelligent communication services for the upcoming 6G \cite{saad2019vision,giordani2020toward}, in supporting the scenarios of seamless and broad-area connection 
 \cite{tao2025generative} and large-scale internet of things (IoT) \cite{lee2021fast,zhou2020deep}.
In such integrated framework, the global navigation satellite system (GNSS) plays a crucial dual role, as not only providing precise timing synchronization required for network coordination, but also delivering centimeter-level positioning services essential for beam alignment in non-terrestrial networks. Therefore, GNSS serves as the spatiotemporal backbone for SAGIN's operations, simultaneously supporting satellite orbit determination in the space layer, unmanned aerial vehicles (UAV) swarm navigation in the air layer, and base station synchronization in the ground layer. On the other hand, with the emerging integrated sensing and communication (ISAC) concept in SAGIN\cite{mao2024uav}, the positioning capabilities provided by the GNSS system have become increasingly vital \cite{dureppagari2023ntn}. 

Nevertheless, as electromagnetic environments become increasingly complex, the integrity of GNSS operations faces escalating risks from interference within its frequency bands, thereby challenging the efficient functioning of SAGIN. Notably, Eurocontrol’s Voluntary Air Traffic Management incident monitoring \cite{EUROCONTROL} has documented an exponential rise in global positioning system (GPS) disruptions over the past five years, with further increases anticipated. A principal source of these disruptions is the proliferation of jamming signals within the GNSS frequency spectrum.
In general, jamming signals are broadly classified into two categories: human-induced and environment-related. Human-induced interferences are either deliberate, such as intentional jamming actions, or accidental, stemming from other electronic systems or intermodulation effects. Conversely, environment-related interferences involve natural phenomena such as multipath propagation, atmospheric scintillation, or signal fading. Deliberate jamming, characterized by its narrowband signal disturbance, poses a notable challenge due to its power levels, which significantly exceed those of GNSS signals. This power discrepancy is primarily due to the proximity of jamming devices to the Earth’s surface, in stark contrast to GNSS signals transmitted from satellites located approximately $19,000$–$23,000$ kilometers above the Earth. Consequently, the path-loss attenuation experienced by GNSS signals far exceeds that affecting jamming signals, creating a pronounced power imbalance that facilitates effective jamming.

To address these challenges, recent research efforts have introduced robust integrity strategies aimed at countering jamming.
These strategies broadly encompass detection, mitigation, localization, and classification of jamming signals.
While detection, mitigation, and localization aspects have seen substantial scholarly attention—documented in studies such as those by \cite{ioannides2016known, morales2019lab, lineswala2019performance} for detection; \cite{ioannides2016known, rezaei2016new, mao2016robust, heng2014gnss} for mitigation; and \cite{amin2016sparse, morales2019lab} for localization—the classification of jamming signals, especially in scenarios involving compound jamming in high jamming-to-noise ratio (JNR) environments, remains relatively underexplored.
Thus, this gap highlights the need for targeted research on jamming classification, which is vital for developing comprehensive jamming management frameworks that can accurately differentiate among various interference types.

In general, the classification delineates five principal jamming types according to their interference patterns and operational characteristics:  (i) Single-tone jamming, as entailing a single continuous frequency that disrupts GNSS signals; (ii) Multi-tone jamming, as employing multiple frequencies concurrently for a more intricate form of interference; (iii) Partial-band noise jamming,  as affecting only a specific segment of the GNSS frequency band; (iv) Linear frequency modulation jamming, distinguished by a frequency that continuously sweeps through a range; and (v) Periodic pulse noise jamming, wherein GNSS signals are intermittently disrupted at regular intervals.
Notably, this study fills the gap in the existing literature by classifying those not only single but also compound jamming interference within GNSS bands under low JNR conditions. This original contribution not only provides a rigorous theoretical framework of GNSS interference but also offers valuable practical insights for jamming detection and mitigation. By devising the classification model that distinguishes between multiple and combined forms of jamming, this work lays a critical foundation for subsequent research and development in GNSS interference management.

The contributions of this paper are summarized as follows:
Firstly, we concentrate on a more realistic portrayal of GNSS jamming in practical situations, highlighting the impact of compound interference on GNSS signals—an aspect that earlier GNSS interference studies have largely overlooked.
Secondly, in response to the challenges associated with identifying compound jamming signals within GNSS signals, we introduce ACSNet–a novel neural network architecture. ACSNet integrates Asymmetric Convolution Blocks (ACB), designed specifically to improve the network's capacity to detect nuanced signal feature variations, thereby improving the recognition of intricate interference patterns.
Thirdly, simulation results confirm ACSNet’s effectiveness in recognizing diverse patterns of compound jamming signals, even at low JNR conditions. Additionally, ACSNet demonstrates notable resilience to variations in the PR of compound jamming signals, highlighting its adaptability to real-world environments where interference signals frequently exhibit imbalanced power distributions.
The remainder of this paper is organized as follows: Section II surveys related works. Section III briefly outlines the mathematical background, covering time-frequency analysis and GNSS signal modeling. Section IV presents the problem formulation, elaborates on the ACSNet design, and describes the simulation dataset. Section V provides the simulation results and comparisons. Finally, Section VI concludes with a summary and reflections on our research.

\section{Related Works}
Recent advancements in the field of GNSS have increasingly leveraged machine learning (ML) techniques to counteract jamming attacks, highlighting ML’s growing dominance in this domain. Specifically, a comparative study in \cite{khoei2024comparative} assessed multiple ML models, emphasizing their effectiveness in detecting and classifying GNSS interference. Notably, decision trees excelled in both classification and regression tasks, outperforming other supervised and unsupervised algorithms.
Further analysis in \cite{shafique2021detecting} explored several ML approaches for jammer detection, concluding that the support vector machine (SVM) exhibits robust detection capabilities. Conversely,  \cite{gallardo2020scer} ranked K-Nearest Neighbors (KNN) above SVM, crediting KNN with superior accuracy in their multi-method comparison.
Additionally, \cite{semanjski2020gnss} corroborated the findings of \cite{shafique2021detecting}, further underscoring SVM’s potential for identifying jamming signals. Overall, these studies collectively underscored the effectiveness of ML in bolstering GNSS resilience against advanced interference methods, while underscoring the importance of aligning specific ML models to particular GNSS jamming detection tasks.

Recent advancements in signal processing and machine learning have catalyzed innovative approaches to the detection and classification of disruptive signals across various communication spectra. Notable among these are the pioneering studies by Swinney and Woods, which employed spectrograms and power spectral density (PSD) metrics. Their research, detailed in \cite{swinney2021unmanned} and \cite{swinney2020unmanned}, leveraged convolutional neural networks (CNNs) to identify and categorize UAVs operating in the 2.4 GHz Wi-Fi bands.
Similarly, Tandiay \emph{et al.} \cite{tandiya2018deep} explored the representation of spectrograms as 2D images within a video prediction framework, specifically employing PredNet \cite{lotter2016deep}, to discern jamming signals.
In a broader application of CNNs, O’Shea \emph{et al.} \cite{o2017spectral} analyzed wireless spectrum data, effectively spotting and classifying a range of signals, including GSM, Bluetooth, and LTE.
Within wireless communications, Arjoune \emph{et al.} \cite{arjoune2020novel} compared SVM, random forest (RF), and neural network models for jamming detection.
Additionally, Wu \emph{et al.} \cite{wu2017jamming} investigated the use of CNNs as feature extractors to classify jamming signals in satellite networks, particularly in complex scenarios containing multiple forms of combined jamming.
Moreover, Junfei \emph{et al.} \cite{junfei2018barrage} and Shao \emph{et al.} \cite{shao2020convolutional} extended jamming detection methods to radar systems, underscoring the breadth of research focused on signal disruption identification in modern communication environments.

In the dynamic landscape of GNSS jamming detection and classification, diverse studies have employed distinct strategies to tackle the complex challenges posed by signal interference. Specifically, Lineswala’s work \cite{lineswala2019performance} offered a thorough exploration of detection methods, leveraging power spectral density analysis to identify jamming signals from the Indian satellite constellation. Despite contributing significantly to GNSS signal detection insights, this study stops short of venturing into jamming signal classification—a step that would yield more detailed information about the nature and source of interference.
Conversely, the research in \cite{morales2019jammer} framed jamming signal classification as an image classification problem by using spectrograms, an innovative approach that views signal spectra as visual data. This strategy facilitates the application of established image processing algorithms to signal analysis. The results are remarkable, with the support vector machine (SVM) model attaining a classification accuracy of 94.90\%, outperforming the convolutional neural network (CNN).

Distinguished from previous research, this work focuses on classifying compound jamming signals within GNSS frequency bands. Additionally, to address this complex task, we propose ACSNet, a novel neural network architecture incorporating asymmetric convolution blocks. This specific design enhances the network’s sensitivity to subtle signal variations, thereby improving classification accuracy. ACSNet’s robustness and accuracy underscore its practical viability in bolstering the security and reliability of GNSS operations. Our results demonstrate ACSNet’s effectiveness in a critical aspect of satellite communication, and point toward further research and applications in signal integrity and interference management.
\section{Mathematical Model}
\subsection{Time-Frequency Analysis Based on CWD}
The sources of GNSS interference can stem from multiple irregular devices or systems, and the amplitude and frequency components of these interference signals may vary over time or space, showing non-stationary characteristics.
Consequently, traditional frequency-based methods, such as the Fourier transform, may fail to capture vital time-domain information and instantaneous frequency shifts\cite{lin2024gnss}. To address this, identifying when each frequency emerges in the signal is essential for time-frequency (TF) analysis. The Choi–Williams Distribution (CWD), a specific TF analysis method, decomposes the signal into sub-signals with distinct frequency components, thereby enabling the extraction of its time-frequency features. Additionally, the CWD introduces a weighting factor that diminishes cross-term interference, retaining accurate time-frequency information. In scenarios involving non-stationary signals or those with intricate features, the CWD surpasses conventional methods like the Short-Time Fourier Transform (STFT) or Wigner–Ville distribution. The CWD’s mathematical definition is expressed as
\begin{equation}\label{eq:1}
W(t,f) = \int_{-\infty}^\infty \int_{-\infty}^\infty A_x(\eta,\tau) \Phi(\eta,\tau) e^{ j2\pi \left (\eta t-\tau f \right )}\mathrm{d} \eta \mathrm{d}\tau,
\end{equation}
where $A_x(\eta,\tau) = \int_{-\infty}^\infty x(t+\frac{\tau}{2}) \overline{x(t-\frac{\tau}{2})} e^{-j2{\pi}t\eta}\mathrm{d}t$, $W(t, f)$ represents the time-frequency distribution at time $t$ and frequency $f$, and $\Phi(\eta,\tau)$ denotes the weighting or window function, designed to mitigate cross-term interference in the time-frequency representation. This function maintains a trade-off between frequency and temporal resolution. Additionally, $\overline{x(t-\frac{\tau}{2})}$ denotes the complex conjugate of $x(t-\frac{\tau}{2})$.
The time-frequency map obtained via CWD effectively represents the spectral distribution of the signal over time, capturing dynamic frequency variations. This capability is particularly advantageous for interference signal identification.

\subsection{GNSS Signal Model} \label{Sec:III.B}
Assume that the signal received at the GNSS receiver is denoted as $r(t)$, adhering to the model
\begin{equation}\label{eq:2}
    r(t) = s(t) + j(t) + n(t),
\end{equation}
where $s(t)$ symbolizes the useful signal, articulated as \cite{borio2017robust}
\begin{equation}\label{eq:3}
    s(t) = \sqrt{2C} d(t-\tau_0) c(t-\tau_0) e^{j{2\pi(f_c+f_0)t +j \phi_0}},
\end{equation}
in which $C$ represents the signal power, $d(\cdot)$ encapsulates the navigation message, $c(\cdot)$ corresponds to the pseudo-random sequence used for transmitting the satellite’s positional data and aiding the receiver in synchronization.
$\tau_0$, $\phi_0$, and $f_0$ denote the time, phase, and Doppler frequency shifts imposed on the received signal due to channel effects, respectively, with $f_c$ indicating the carrier center frequency.
Additionally, $n(t)$ is characterized as additive white gaussian noise (AWGN), and $j(t)$ represents the possible jamming signals, 
where compound jamming signals may be engendered through diverse operations such as addition, convolution, or multiplication.
We specifically address nine types of compound jamming in GNSS, constructed through the pairwise superposition of five prevalent jamming signals \cite{lin2024gnss,morales2019jammer} within the GNSS system in the time domain. The expressions for these five jamming signals are delineated subsequently,

{\em 1) Single-Tone Jamming (STJ): } STJ is characterized by a single impulse in the frequency domain, and its power-normalized model is given as 
\begin{equation}\label{eq:4}
    j_1(t) = e^{j{(2\pi f_c t + \phi)}}.
\end{equation}
Here, \( f_c \) specifies the center frequency, and \( \phi \) represents the initial phase of the interference signal.

{\em 2) Multi-Tone Jamming (MTJ): } MTJ is characterized by the presence of multiple independent, non-overlapping impulse components in the frequency domain, and its power-normalized model is given as 
\begin{equation}\label{eq:5}
    j_2(t) = \frac{1}{A}\sum_{k=1}^{K}\sqrt{P_k} e^{j{(2\pi f_k t + \phi_k)}},
\end{equation}
where $A=\sqrt{\sum_{k=1}^{K}{P_k}}$, $P_k$ denotes the signal power of the $k$-th tone, and $K$ represents the total number of tones, encapsulating the complexity and breadth of the jamming signal’s spectrum.

{\em 3) Partial-Band Noise Jamming (PBNJ): } PBNJ is characterized by its selective interference within a specific segment of the signal’s frequency band. In particular, this type of jamming impacts only certain frequency ranges, leaving other parts of the band unaffected by jamming signals, and its power-normalized model is given as 
\begin{equation}\label{eq:6}
    j_3(t) = a u(t) e^{j{(2\pi f_c t + \phi)}},
\end{equation}
where $a$ is the power normalization factor, and \( u(t) \) represents a band-limited Gaussian white noise signal with the mean of $0$ and the variance of ${\sigma_n}^2$.

{\em 4) Linear Frequency Modulation Jamming (LFMJ): } LFMJ is a specific type of frequency modulation interference where the frequency of the jamming signal is modulated linearly as varying continuously within a predetermined frequency range, and its power-normalized model is given as
\begin{align}
j_4(t) &= e^{j(2\pi f_c t + \pi q t^2 + \phi)}, \quad 0 \leq t \leq T,\label{add:7} \\
B &= qT,\label{add:7-1}
\end{align}
where \( q \) indicates the rate of frequency change over time, encapsulating the linear nature of the frequency modulation.
Additionally, in Eq. \eqref{add:7-1}, \( B \) signifies the sweep bandwidth, representing the range of frequency variation, while \( T \) represents the sweep period, as the duration required to complete one full frequency cycle. 

{\em 5) Periodic Pulse Noise Jamming (PPNJ): } PPNJ involves  periodic emission of high-intensity jamming pulses, which disrupt the normal operation of the target receiver by instantaneously overwhelming its processing capabilities. The power-normalized model of PPNJ is given as 
\begin{equation}\label{eq:8}
j_5(t)=\begin{cases}
\sqrt{\frac{T}{\tau}}, &|t| \leq \frac{\tau}{2},\\
0, &otherwise,
\end{cases}
\end{equation}
where \( T \) specifies the pulse period, \( \tau \) denotes the duration of the square wave pulse within each period. 

Considering a compound jamming signal $j(t)$ formed by the superposition of two interference signals,  \( j_i(t) \) and \( j_k(t) \), whose expressions are given above, $j(t)$ is then computed as
\begin{equation}\label{eq:9}
    j(t) = \alpha j_i(t) + \beta j_k(t), \quad 
    i, k \in \{1, \dots, 5\}, \quad i < k,
\end{equation}
where $\alpha$ and $\beta$ denote the power scaling factors of $j_i(t)$ and $j_k(t)$, respectively.
Furthermore, the compound jamming power ratio, PR, can be defined as
\begin{equation}\label{eq:10}
    \text{PR} = \frac{\|\beta\|^2}{\|\alpha\|^2}.
\end{equation}

\begin{figure}
\centering
\includegraphics[width=0.4\textwidth]{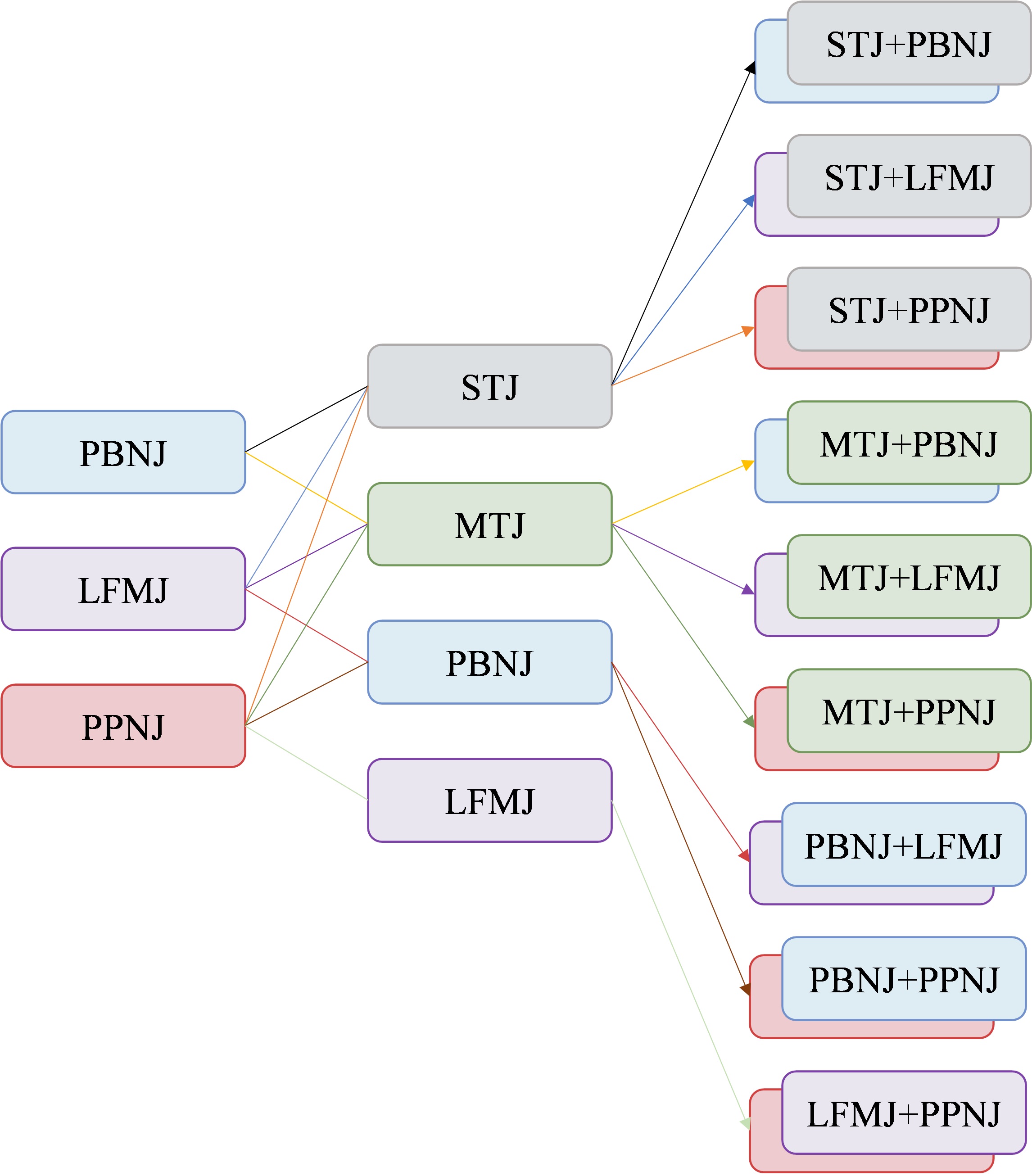}
\caption{The hierarchical structure of different compound interferences. The diagram illustrates the formation of compound jamming types by combining basic interference signals, including STJ, MTJ, PBNJ, LFMJ, and PPNJ.}
\label{fig:11}
\end{figure}
\begin{figure*}
\centering
\includegraphics[width=1.0\textwidth]{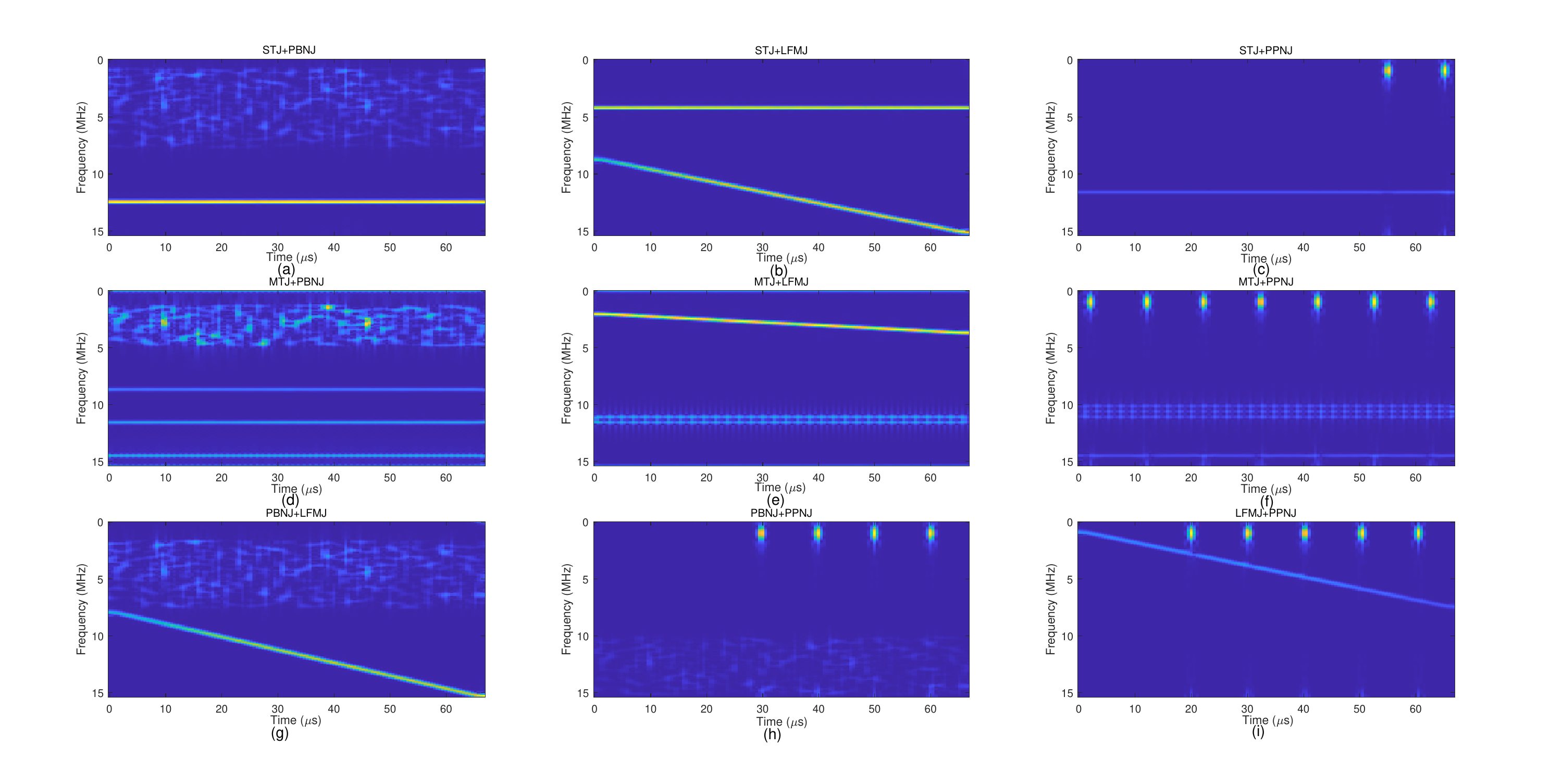}
\caption{The time-frequency spectrograms of the compound GNSS jamming:
(a) STJ+PBNJ, (b) STJ+LFMJ, (c) STJ+PPNJ, (d) MTJ+PBNJ, (e) MTJ+LFMJ, (f) MTJ+PPNJ, 
(g) PBNJ+LFMJ, (h) PBNJ+PPNJ, (i) LFMJ+PPNJ.}
\label{fig:1}
\end{figure*}

Since the combination of single-tone and multi-tone jamming behaves like multi-tone jamming alone, we focus on the remaining nine possible interference combinations, as shown in Fig. \ref{fig:11}. Additionally, to visually articulate the impact of these combinations on the signal’s spectral characteristics, time-frequency spectrograms of these nine compound jamming signals are meticulously presented in Fig. \ref{fig:1}.
Here, the carrier-to-noise ratio $C/N_0$ is considered to be a fundamental measure in communication systems, quantifying the ratio of the carrier power of the received signal $C$ to the spectral density of noise power $N_0$.
As such, it serves as a key indicator of signal robustness against background noise.

Additionally, JNR is considered as a key metric for assessing the presence and intensity of interference relative to noise \cite{lin2024gnss},
and defined as 
\begin{equation}\label{eq:11}
    \text{JNR} = 10 \log_{10} \left( \frac{P_J}{P_N} \right),
\end{equation}
where $P_J$ denotes the interference signal power, and \( P_N \) represents the noise power. 

After sampling \( j(t) \), the discrete sequence \( j[n] \) is obtained. Given \( N \) sampling points, the average power of the interference signal is computed as
\begin{equation}\label{eq:12}
    P_J=\frac{1}{N}\sum_{n=1}^{N}|j[n]|^2.
\end{equation}

In the context of normalization, the corresponding mathematical formulation is given as
\begin{equation}\label{eq:13}
    y[n]=\frac{x[n]}{\sqrt{\sum_{n=1}^{N}|x[n]|^2}},
\end{equation}
where $x[n]$ represents the original signal samples, and $y[n]$ denotes the normalized signal samples.

\section{Methodology}
\subsection{Problem Formulation}
Assuming the GNSS receiver captures \( N \) sample signals \( r(t) \), their respective time-frequency spectrograms \( x^{(n)} \) are obtained using \eqref{eq:1}. The input space is then defined as a set of pairs, \( D = \{(x^{(n)}, y^{(n)})\}_{n=1}^{N} \), where \( x^{(n)} \) represents the time-frequency spectrogram of the \( n \)-th jamming signal, and \( y^{(n)} \) denotes the corresponding label.
Building on the methodology outlined in \cite{qu2020jrnet}, label \( y^{(n)} \) is obtained by applying the mapping function \( g(x^{(n)}) \) to the input \( x^{(n)} \). The primary objective is to develop a predictive model \( f(x; \theta) \) that closely approximates the true mapping function \( g(x^{(n)}) \).
To achieve this, we optimize the model parameters \( \theta \) to ensure that the model output \( f(x; \theta) \) closely matches the actual labels \( y^{(n)} \), achieved by minimizing a loss function that quantifies the discrepancy between predicted and true labels, thereby improving the model’s accuracy and generalization. The predictive model \( f(x; \theta) \) is mathematically defined as
\begin{equation}\label{eq:14}
    f(x; \theta) = w^T \phi(x) + b,
\end{equation}
where \( \phi(x) \) denotes the feature vector, \( w \) represents the weight vector, and \( b \) is the bias term.

The risk function \( R(\theta) \) evaluates the overall performance of the model especially when dealing with large datasets, defined as
\begin{equation}\label{eq:15}
    R(\theta) = \mathbb{E}_{(x, y) \sim p(x, y)} \left[ L(y, f(x; \theta)) \right],
\end{equation}
where $L(y, f(x; \theta))$ is the loss function, and \( p(x, y) \) represents the joint distribution of inputs \( x \) and labels \( y \).
In this study, we utilize the cross-entropy loss function, specifically designed for classification tasks as
\begin{equation}\label{eq:16}
    L(y, f(x; \theta)) = - \sum_{v=1}^{V} y_v \log(f(x; \theta)_v),
\end{equation}
where $V$ denotes the number of classes, $y_v$ is the one-hot encoded label for class $v$, and $f(x; \theta)_v$ indicates the predicted probability for class $v$.
The primary objective of model training is to minimize the expected cross-entropy loss \( R(\theta) \), formulated as
\begin{equation}\label{eq:17}
    \theta^* = \arg \min_{\theta} R(\theta).
\end{equation}

Additionally, we employ the adaptive moment estimation (Adam) optimizer for its efficiency and robustness in handling large-scale optimization problems \cite{kingma2014adam}.
It enhances the conventional gradient descent process by incorporating momentum and scaling the learning rates for each parameter independently based on historical gradient information.
This is achieved through the computation of the first and second moment estimates of the gradients, which are updated at each iteration as follows.

For the first moment estimate, or the exponentially weighted average of past gradients, we calculate
\begin{equation}\label{eq:18}
    m_t = \beta_1 \cdot m_{t-1} + (1 - \beta_1) \cdot g_t,
\end{equation}
where \(m_t\) denotes the first moment estimate at iteration $t$, \(m_{t-1}\) is the estimate from the previous iteration \((t-1)\), \(g_t\) represents the gradient at iteration $t$, and \(\beta_1\) is the decay factor for the first moment components, typically being set to $0.9$.
Similarly, the second moment estimate, or the exponentially weighted average of squared gradients, is updated as
\begin{equation}\label{eq:19}
    v_t = \beta_2 \cdot v_{t-1} + (1 - \beta_2) \cdot g_t^2,
\end{equation}
where \(v_t\) is the second moment estimate at iteration $t$, \(v_{t-1}\) is the estimate from the previous iteration, and \(\beta_2\) is the decay rate for the second moment components, generally being set to $0.999$.

\subsection{Overview of traditional CNNs}
CNNs have gained widespread recognition as one of the most effective tools for image recognition tasks, driven by their unique architectural attributes and sophisticated training mechanisms.
At the core of CNNs’ functionality lies in their ability to perform convolutional operations that automatically learn and extract significant features from images. 
This capability crucially eliminates the need for manual feature extraction, a process traditionally required in earlier image processing methods.
One of the distinct advantages of CNNs is their method of feature extraction \cite{varshni2019pneumonia,barbhuiya2021cnn}, which is inherently adaptive.
Further enhancing their capability, CNNs integrate a hierarchical feature learning mechanism.
Specifically, this mechanism, combined with local receptive fields, convolution operations, and pooling layers, enables the networks to construct a profound comprehension of the image’s spatial hierarchy.
As the network progresses, through successive layers, CNNs can capture low-level features such as edges and textures in the initial layers, and progressively higher-level features in deeper layers.
This structured approach to feature learning is particularly beneficial for capturing the nuanced patterns and variations in images associated with different types of jamming signals.

Specifically, the convolutional layer serves as a foundational component of CNNs, playing a pivotal role in the automatic extraction of features from the input data.
This layer utilizes a convolution operation that involves sliding a kernel across the input image.
At each step of this sliding motion, the kernel performs a dot product operation with a local region of the input image, effectively capturing local patterns and features, such as edges, textures, or other fundamental visual elements.
As a result, these operations produce a new feature map that represents these local features at a higher abstraction level than the raw input.
As described in \cite{qu2020jrnet}, the mathematical representation of the output for a neuron in the $l$-th convolutional layer of a CNN can be expressed by 
\begin{equation}\label{eq:20}
    z_{i,j,k} = \sum_{u=1}^{f_{h'}} \sum_{v=1}^{f_{w'}} \sum_{k' = 1}^{f_{n'}} x_{i',j',k'} \cdot w_{u,v,k',k} + b_k.
\end{equation}
where \(z_{i,j,k}\) denotes the output of the neuron at position $(i,j)$ in the $k$-th feature map of the $l$-th convolutional layer,
$x_{i',j',k'}$  corresponds to the output from the previous layer’s feature map at position $({i',j',k'})$,
and $w_{u,v,k',k}$ represents the weights of the convolutional kernel, where $u$ and $v$ index the spatial dimensions (height and width, respectively) of the kernel.
Additionally, $b_k$ is added to the output of the convolution operation for the $k$-th feature map, enhancing the neuron’s ability to fit the data correctly,
while dimensions $(f_{h'},f_{w'},f_{n'})$ refer to the height, width, and depth of the feature map from the preceding layer.
\subsection{The construction of ACSNet}
\begin{figure}
\centering
\includegraphics[width=0.35\textwidth]{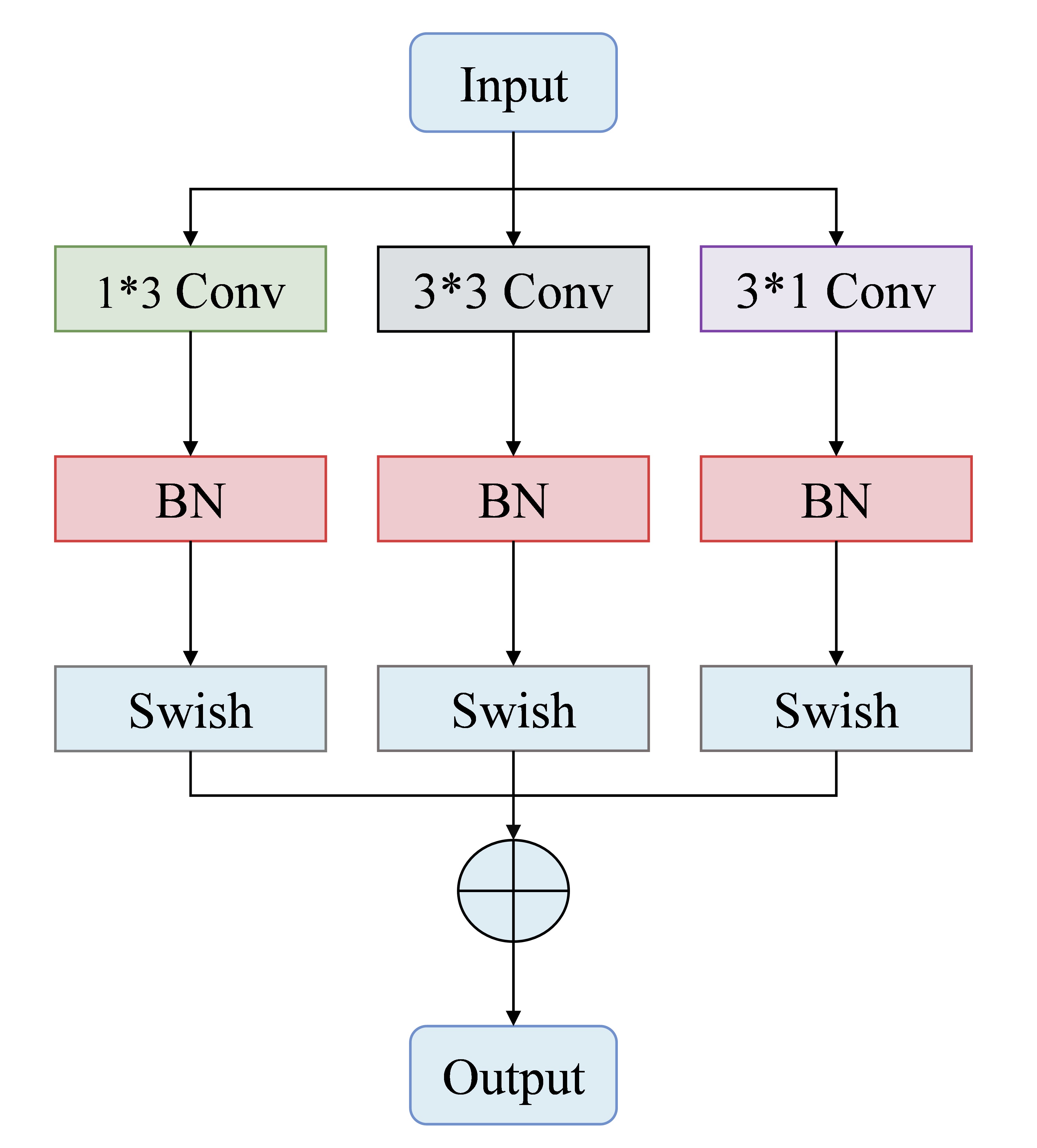}
\caption{The structure of ACB.}
\label{fig:2}
\end{figure}

To enhance model efficiency, particularly in environments constrained by computational resources or lacking robust hardware acceleration, the utilization of ACB \cite{ding2019acnet} is considered.
ACBs represent a significant departure from traditional convolutional models that typically employ square convolution kernels.
By replacing these larger kernels with combinations of smaller and rectangular ones, such as $1\times3$ and $3\times1$, ACBs achieve a substantial reduction in the number of parameters within the convolution kernels.
This innovative approach offers a dual benefit as it markedly diminishes the computational complexity and concurrently reduces memory usage, thereby elevating the overall computational efficiency of the network.
The architectural design of ACB employed in this study is illustrated in Fig.\ref{fig:2}. Specifically, each ACB is composed of three parallel convolution layers, each characterized by a distinct kernel size of $1\times3$, $3\times3$, and $3\times1$.
This configuration allows the network to capture and integrate features from different directional perspectives, significantly enhancing its representational capabilities.
To optimize the performance and stability of the network, the output from each convolution layer within the ACB undergoes batch normalization (BN).
This step is crucial as it facilitates faster network convergence and mitigates the risk of overfitting by standardizing the outputs before they proceed to the next layer.
Subsequent to the normalization, the outputs are subjected to a non-linear transformation through an activation function, which further enhances the network’s capacity to model complex patterns inherent in the data.

The overall architecture of the ACSNet integrates six such ACBs, each contributing to a more granular and robust feature extraction process.
This structure is specifically tailored to maximize efficiency and effectiveness in feature learning, ensuring that the network remains computationally viable even in settings with limited processing power.
BN plays a pivotal role in enhancing the training dynamics of CNNs by normalizing the input layer data distributions.
This normalization process is crucial for stabilizing the learning process, as it effectively mitigates common issues such as vanishing or exploding gradients.
By ensuring a more consistent distribution of inputs across the network layers, BN allows for the utilization of higher learning rates, which in turn accelerates the convergence speed of the network training.
Consequently, BN addresses and alleviates many of the training bottlenecks that are frequently encountered in traditional neural network training methods.
Following the BN process and a subsequent linear scaling transformation, the output \(z_{i,j,k}\) from \eqref{eq:20} undergoes further transformation as
\begin{equation}\label{eq:21}
O_{:,:,j} = \frac{\sum_{k=1}^{C} M_{:,:,k} * F^{(j)}_{:,:,k} - \mu_j}{\sigma_j} \gamma_j + \beta_j
\end{equation}
where $O_{:,:,j}$ denotes the output after BN and linear scaling for the $j$-th channel.
Here, $\mu_j$ and $\sigma_j$ represent the mean and standard deviation of the $j$-th channel, respectively.
Additionally, $\gamma_j$ and $\beta_j$ are learnable parameters that scale and shift the normalized data, respectively, enhancing the network’s ability to fit the specific data characteristics.

According to the framework established in \cite{ding2019acnet}, the overall output after BN within an ACB can be calculated as
\begin{equation}\label{eq:22}
O_{ACB} = O_{:,:,j} + \overline{O}_{:,:,j} + \hat{O}_{:,:,j}
\end{equation}
where $O_{:,:,j}$, $\overline{O}_{:,:,j}$ and $\hat{O}_{:,:,j}$ represent the outputs from the $3\times3$, $1\times3$ and $3\times1$ convolutional layers, respectively.
\begin{figure*}
\centering
\includegraphics[width=1.0\textwidth]{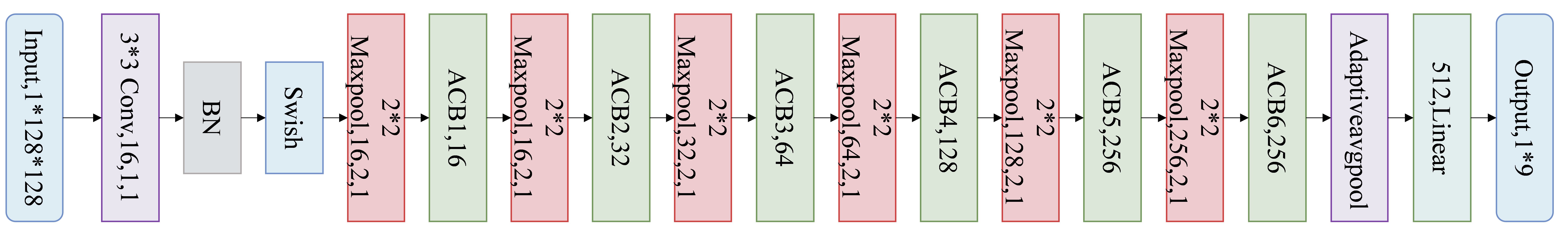}
\caption{The structure of ACSNet: ``$3*3$ Conv, $16,1,1$'' means a $3*3$ convolutional kernel, $16$ filters, $1$ stride and $1$ padding, ``$2*2$ Maxpool, $16,2,1$'' represents a $2*2$ maxpool layer, $16$ channels, $2$ strides and $1$ padding, ``ACB1, $16$'' represents an ACB shown in Fig.\ref{fig:6} with $16$ channels, ``Adaptiveavgpool'' means a global average pool with a $(1*1)$ output, and ``$512$, Linear'' means a fully connected layer.}
\label{fig:3}
\end{figure*}
In the design of ACSNet, the Swish function \cite{ramachandran2017searching} is selected as the activation function to facilitate the introduction of non-linear features into the network architecture.
This choice is driven by the Swish function's several distinctive properties that contribute to its effectiveness in deep learning applications. 
Specifically, these properties include its smoothness, non-monotonicity, a bounded lower limit, and an unbounded upper limit, each contributing uniquely to the function’s performance in neural networks \cite{nwankpa2018activation}.

One of the most significant advantages of the Swish function lies in its smoothness.
In particular, this characteristic allows for a more continuous and differentiable function at all points, beneficial for optimization algorithms that rely on gradient-based methods.
As a result, the smooth nature of Swish helps in achieving better optimization landscapes, thus facilitating more efficient training processes \cite{ramachandran2017searching}.
Moreover, the non-monotonic behavior of the Swish function allows it to preserve and propagate negative information, which can be crucial for certain types of data, such as signals with bipolar characteristics, financial time series with fluctuations, or datasets where negative values carry meaningful semantic information (e.g., temperature anomalies or sentiment scores). By maintaining these negative values instead of suppressing them, Swish can help neural network models capture more nuanced patterns and improve generalization in tasks like time-series forecasting, natural language processing, and signal processing.
The corresponding mathematical expressions is shown as
\begin{equation}\label{eq:23}
f(x) = x \cdot sigmoid(x) = \frac{x}{1+e^{-x}}.
\end{equation}

Moreover, the Swish activation function is distinguished by its continuous and differentiable nature, which significantly mitigates issues related to gradient explosions commonly encountered with other activation functions.
Unlike the traditional rectified linear unit (ReLU) function, Swish facilitates more efficient gradient propagation throughout the network.

Additionally, the pooling layer serves as a vital component within CNNs, primarily utilized to reduce the spatial dimensions of feature maps while preserving essential information. 
By doing so, pooling layers decrease the computational complexity of the model, reduce the number of parameters, and prevent overfitting.
These layers also improve the model’s robustness against input transformations such as translation and rotation, thereby enabling more effective adaptation to varying input data.
Among the common types, pooling methods implemented include Max Pooling, Average Pooling, and Global Average Pooling.

In the architecture of ACSNet, a sophisticated approach combines Max Pooling and Global Average Pooling to optimize feature extraction and network efficiency.
Specifically, Max Pooling is strategically placed following the convolutional layers to diminish spatial dimensions and accentuate significant local features.
Conversely, Global Average Pooling, positioned at the terminal section of the network, compacts the entire feature map of each channel into a singular value.
This operation serves to abstract global features and reduce dimensionality, which not only decreases the parameter count but also fosters the model’s ability to generalize, effectively curbing overfitting.
As a result, this dual pooling strategy ensures that ACSNet retains crucial information while minimizing computational demands and enhancing overall robustness.

Lastly, the structure of ACSNet, depicted in Fig.\ref{fig:3}, epitomizes a robust framework for interference recognition and classification.
The initial stage involves utilizing the CWD to generate a detailed time-frequency map of the GNSS signal, typically resulting in dimensions of $1\times128\times128$.
This map is subsequently processed by ACSNet, which, through its deep learning architecture, extracts pertinent features and performs classification to determine the type of interference, including various compound interferences. 
As a result, this streamlined process not only recognizes different interference types with high efficiency but also delivers precise classification outcomes, even in environments characterized by compound jamming signals.

\subsection{Data processing}
In alignment with the signal model outlined in Section \ref{Sec:III.B}, the GNSS received signal $r(t)$ is generated to operate within the GPS L1 frequency band, characterized by a carrier frequency of approximately $1.57542$ GHz.
The specific signal model is expressed in \eqref{eq:3}, where $C$ is set to $1$.
\( \tau \), \( \phi \), and \( f_0 \) are set as $0$, \( \pi/4 \), and $1000$ Hz, respectively, to simulate typical operational parameters.
To mimic the navigation information carried within the signal, the navigation data \( d(\cdot) \) is generated from a random bit stream.
Concurrently, the pseudo-random noise (PRN) sequence \( c(\cdot) \) employs the C/A code with a PRN number of $1$, reflecting common settings in GNSS simulations.
Additionally, the $C/N_0$ is maintained at $40$ dB$\cdot$Hz to ensure realistic signal strength in relation to background noise.
In this context, the interference signal $j(t)$ encompasses nine types of compound jamming signals as detailed earlier in Section \ref{Sec:III.B}. The essential parameters governing these interference signals, such as the carrier frequency $f_c$, sampling frequency $f_s$, the number of sampling points $N$, bandwidth and phase of the interference signal, are methodically summarized in Table \ref{Table:1}.
Notably, the phase of each interference signal is randomized, following a uniform distribution between $0$ and $2\pi$, to replicate varying interference conditions realistically.
Given the challenge posed by low JNR conditions in signal detection and identification, our study specifically focuses on a dataset with JNR values ranging from $-20$ dB to $10$ dB.
This range is selected to rigorously evaluate the model’s performance in scenarios where the jamming signal is even significantly weaker relative to the noise.
Such conditions, therefore, are particularly challenging to test the model’s ability to effectively detect and classify interference amidst substantial background noise.

In the final stage of the signal processing workflow, the received signal undergoes CWD, which is subsequently refined through a Hamming window to produce its time-frequency spectrogram.
For each type of interference outlined in the dataset, a total of $500$ time-frequency spectrograms are generated at each JNR level.
This extensive collection of data points provides a robust foundation for a comprehensive analytical assessment.
Moreover, to emulate more realistic interference conditions commonly encountered in real-world scenarios, the study extends its analysis to include a wide range of interference power ratios.
These ratios span from $-15$ dB to $25$ dB, encompassing a broad spectrum of potential signal environments—from weak to strong interference intensities.
By evaluating the model’s performance across these varied conditions, the research aims to thoroughly examine its resilience and adaptability.
This rigorous testing across different interference intensities and signal quality levels allows for a detailed assessment of the model’s robustness.
Not only does this approach highlight the model’s capability to adapt to different levels of interference, but it also provides critical insights that are instrumental in optimizing the system’s design.
Such insights, therefore, are invaluable in ensuring that the system maintains optimal efficiency and reliability, even within complex and dynamically changing environments.
\begin{table}[h]
\caption{Jamming Signal and Main Parameters} \label{Table:1}
\centering
\begin{tabular}{ccc}
\toprule  
Jamming signal types & Parameters & Value range \\ \midrule  
\multirow{6}{*}{STJ}& $f_c$ (MHz) & $0$ - $15.36$\\
 & JNR (dB) & $-20$ - $10$\\ &Bandwidth (MHz) & - \\ &$f_s$ (MHz) & $15.36$ \\ &N &$1024$ \\ &Phase&($0,2\pi$) \ \\ 
\midrule
\multirow{6}{*}{MTJ}& $f_c$ (MHz) & $0$ - $15.36$\\
 & JNR (dB) & $-20$ - $10$\\ &Bandwidth (MHz) & - \\ &$f_s$ (MHz) & $15.36$ \\ &N &$1024$ \\ &Phase&$(0,2\pi)$ \ \\ 
\midrule
\multirow{6}{*}{PBNJ}& $f_c$ (MHz) & $0$ - $15.36$\\
 & JNR (dB) & $-20$ - $10$\\ &Bandwidth (MHz) & $1.536$ - $7.68$ \\ &$f_s$ (MHz) & $15.36$ \\ &N &$1024$ \\ &Phase&$(0,2\pi)$ \ \\  
\midrule
\multirow{6}{*}{LFMJ}& $f_c$ (MHz) & $0$ - $15.36$\\
 & JNR (dB) & $-20$ - $10$\\ &Bandwidth (MHz) &$1.536$ - $7.68$  \\ &$f_s$ (MHz) & $15.36$ \\ &N &$1024$ \\ &Phase&$(0,2\pi)$ \ \\ 
\midrule
\multirow{6}{*}{PPNJ}& $f_c$ (MHz) & $0$ - $15.36$\\
 & JNR (dB) & $-20$ - $10$\\ &Bandwidth (MHz) & $5$ \\ &$f_s$ (MHz) & $15.36$ \\ &N &$1024$ \\ &Phase&$(0,2\pi)$ \ \\ 
 \midrule
 \multirow{6}{*}{Compound Jamming}&$f_c$ (MHz)&\multirow{6}{3.3cm}{The same parameter settings as the corresponding single jamming signal}\\
 &JNR (dB)&\\&Bandwidth (MHz)&\\&$f_s$ (MHz)&\\&N&\\&Phase&\ \\
\bottomrule  
\end{tabular}
\end{table}

\section{Simulation Results}
\subsection{Simulation Settings}
In this study, the simulation experiments are structured to optimize both the computational efficiency and the efficacy of the model training.
The dataset is partitioned with an $80$-$20$ split, allocating $80\%$ for training and the remaining $20\%$ for testing.
This allocation ensures a substantial training set while providing an adequate testing set to evaluate the model’s performance. 
To streamline the training process and conserve computational resources, a validation set is not utilized. The batch size for training is established at $64$, and a learning rate of $0.01$ is applied.
The model is subjected to a comprehensive training regimen spanning $400$ epochs, a strategy designed to strike a balance between computational efficiency and the potential for robust model performance.

For the evaluation of the model’s recognition capabilities, two principal metrics are employed: overall accuracy (OA) and the Kappa coefficient (Ka), as detailed in \cite{qu2020jrnet}.
Together, these metrics are instrumental in providing a comprehensive assessment of the model’s classification accuracy,
Specifically, OA quantifies the general accuracy of the model and is computed as 
\begin{equation}\label{eq:24}
OA=\frac{\text{The number of correct samples}}{M},
\end{equation}
where $M$ is the number of samples in the test set.
Meanwhile, Ka measures the agreement between the observed classifications and those that would occur by chance, thereby providing a more nuanced understanding of the model’s performance. The computation of the Kappa coefficient is outlined as
\begin{equation}\label{eq:25}
\begin{cases}
p_e = \frac{a_1 \times b_1 + a_2 \times b_2 + \cdots + a_{9} \times b_{9}}{M^2},\\
Ka=\frac{p_{OA}-p_e}{1-p_e},
\end{cases}
\end{equation}
where $a_i$ represents the number of samples of each type of interference signal in the test set, and $b_i$ denotes the number of samples correctly classified for each corresponding type.
Additionally, the term $p_{OA}$ signifies the probability of overall accuracy.
Together, the values of OA and Ka serve as critical indicators of the model’s effectiveness, with higher values signifying superior performance.

\subsection{Recognition Performance of ACSNet}
\begin{figure*}[ht]
\centering
\subfigure[]{
\begin{minipage}[t]{0.47\linewidth}
\includegraphics[width=3.5in]{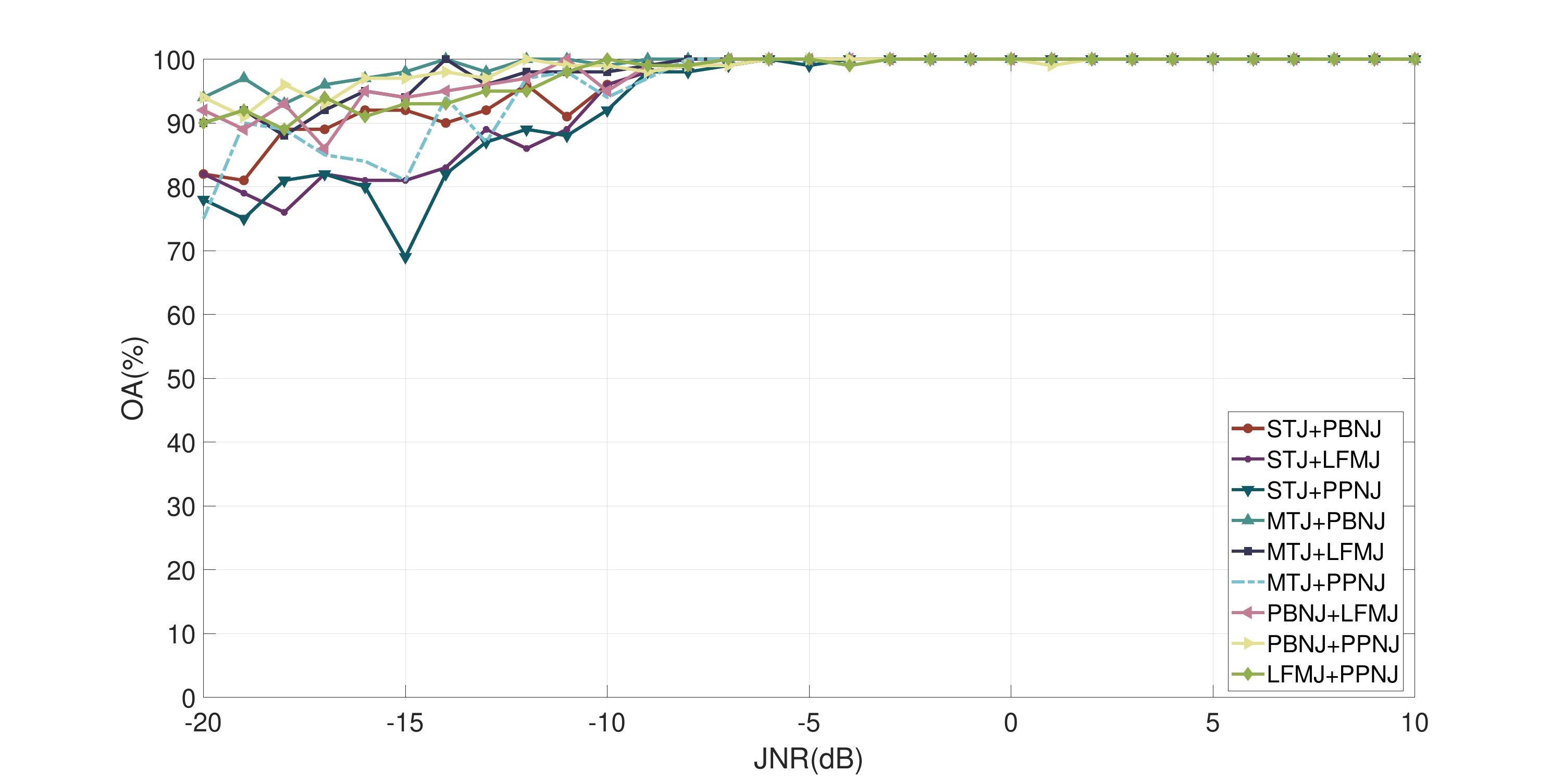}
\end{minipage}
\label{fig:4-1}
}
\subfigure[]{
\begin{minipage}[t]{0.47\linewidth}
\includegraphics[width=3.5in]{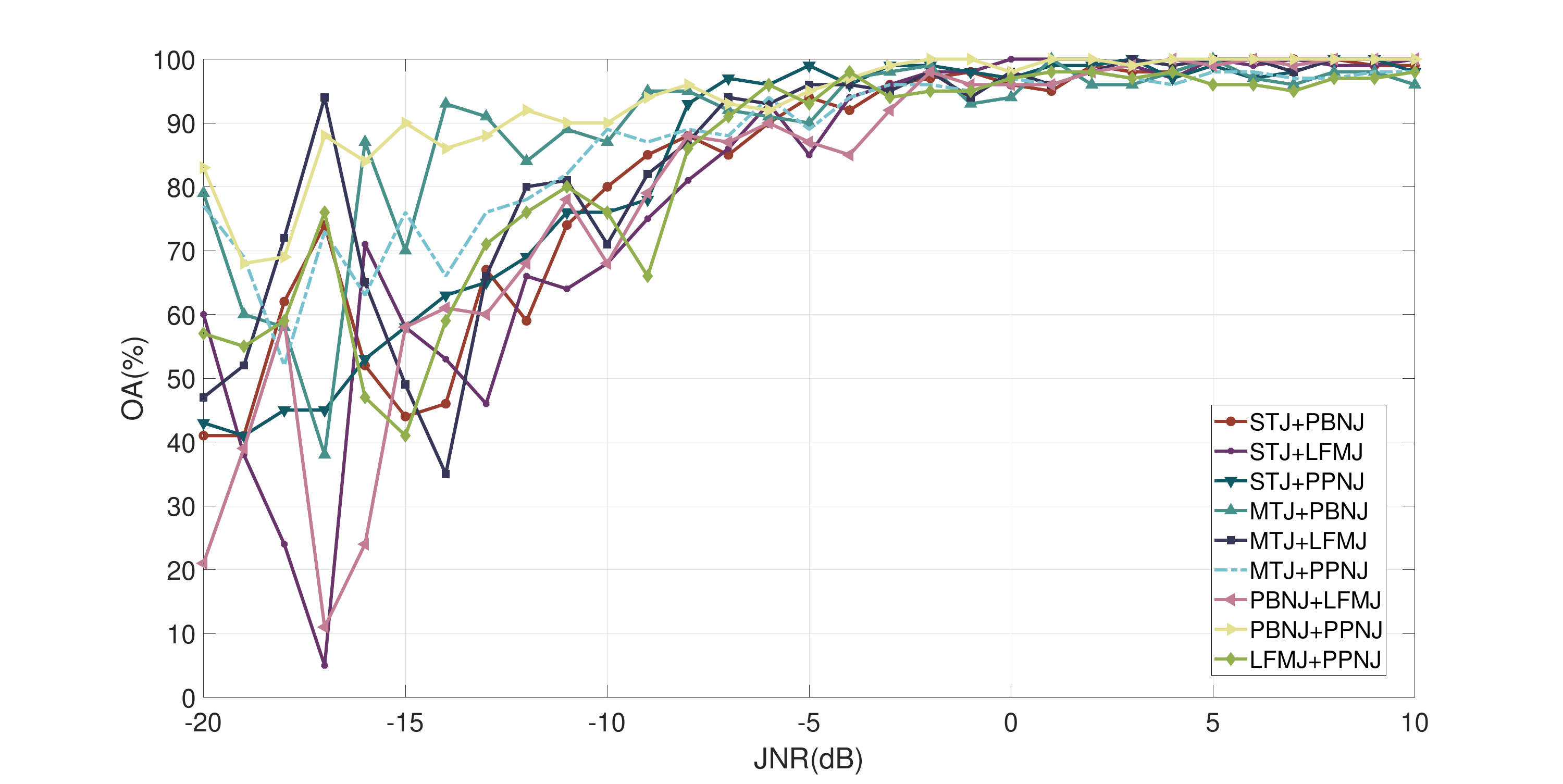}
\end{minipage}
\label{fig:4-2}
}
\subfigure[]{
\begin{minipage}[t]{0.47\linewidth}
\includegraphics[width=3.5in]{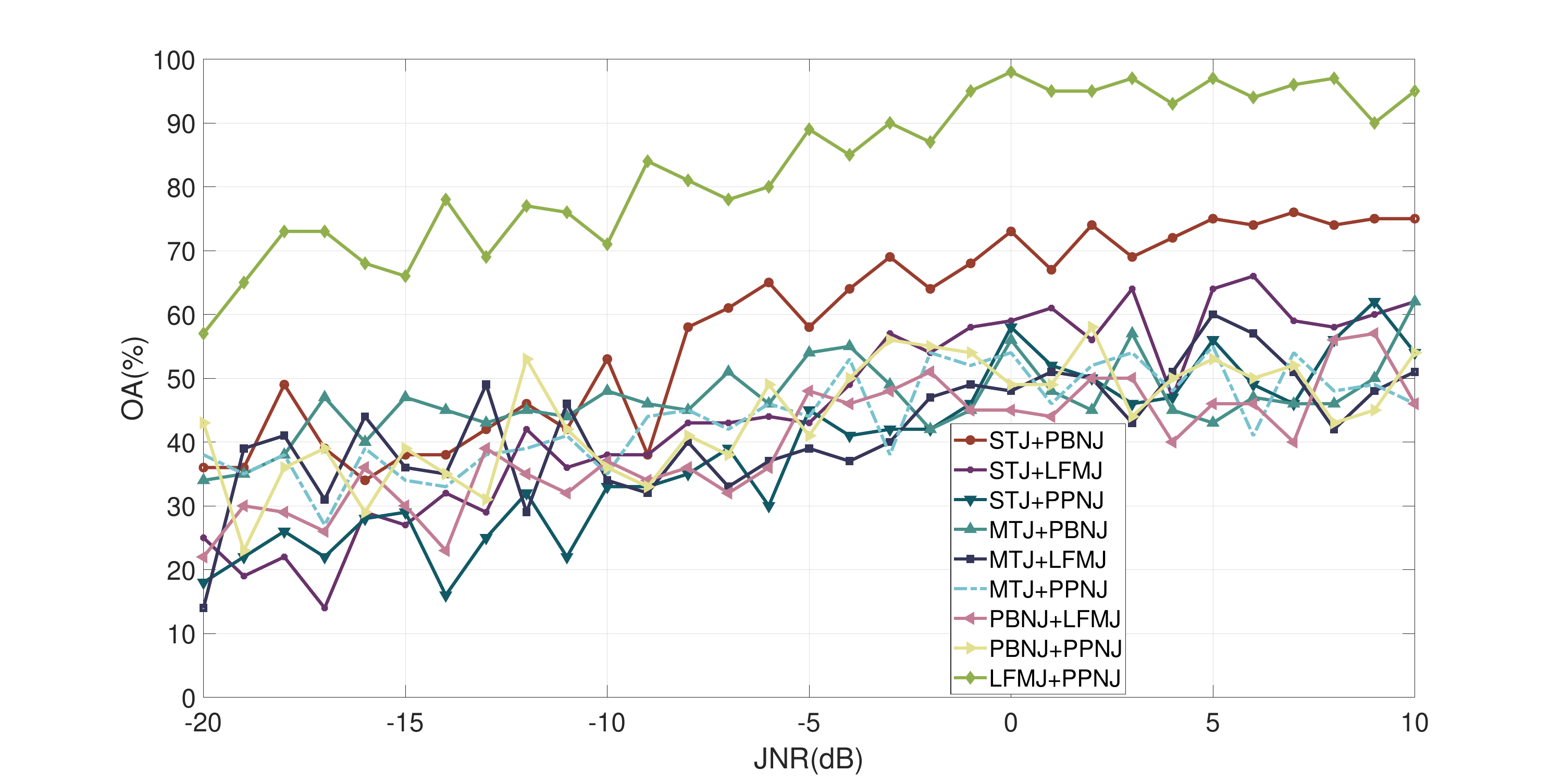}
\end{minipage}
\label{fig:4-3}
}
\subfigure[]{
\begin{minipage}[t]{0.47\linewidth}
\includegraphics[width=3.5in]{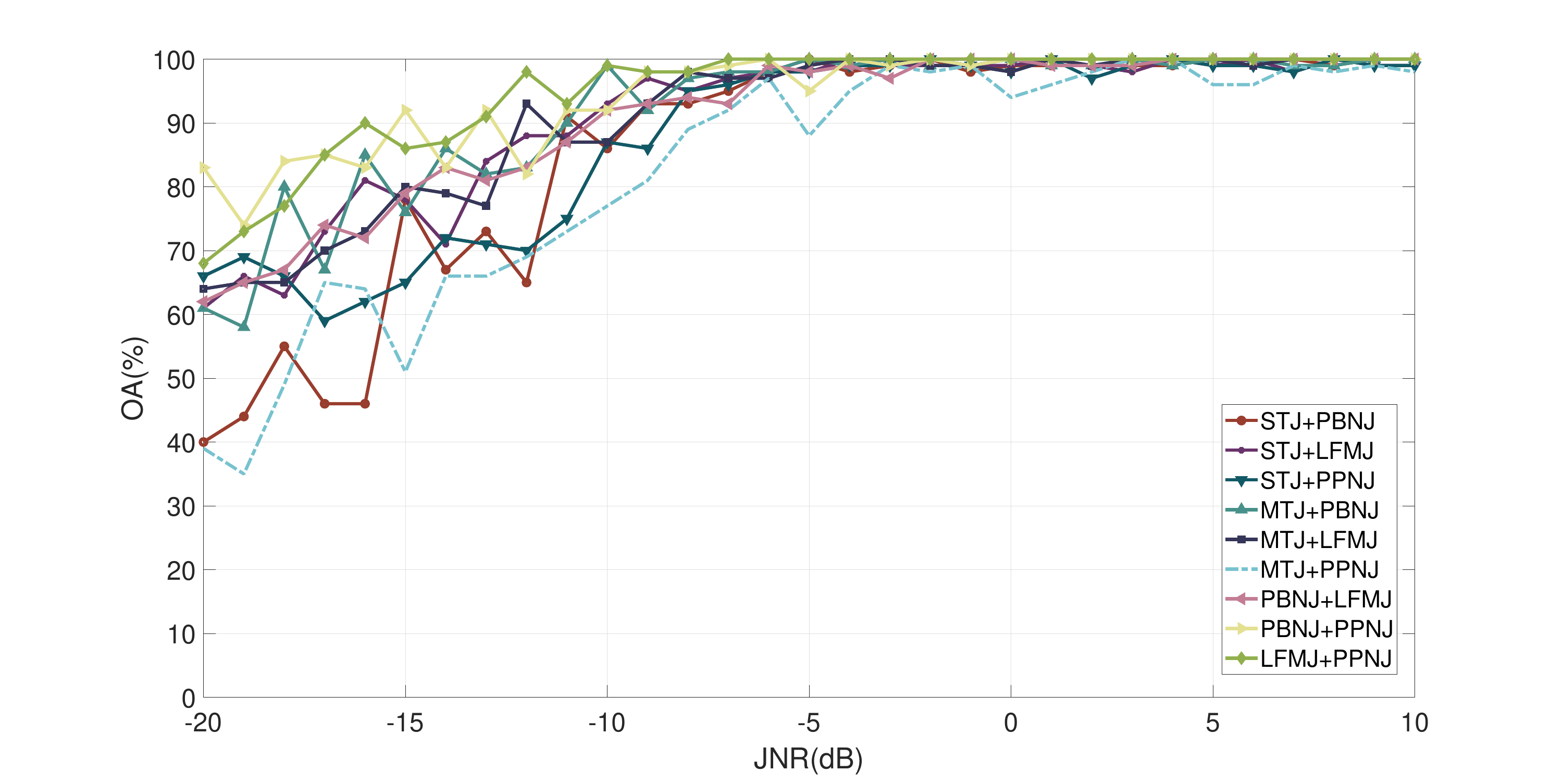}
\end{minipage}
\label{fig:4-4}
}
\caption{The recognition performance at different JNRs of different CNN models:(a) ACSNet, (b) BCNN, (c) FCNN, and (d) RCNN.}
\vspace{-5mm}
\end{figure*}

The effectiveness of the proposed ACSNet in recognizing different types of compound jamming signals under various JNR conditions is depicted in Fig. \ref{fig:4-1}. The OA for each type of compound interference demonstrates the robustness of the model across a spectrum of interference scenarios. Specifically, the OA for MTJ and PBNJ compound jamming, as well as PBNJ and PPNJ compound jamming, consistently exceeds $90\%$ across a JNR range from $-20$ dB to $10$ dB. This high level of accuracy highlights the model’s capability to effectively classify these types of interference under significant noise conditions.
Further results in Fig. \ref{fig:4-1} reveals that the OAs for MTJ and LFMJ compound jamming, PBNJ and LFMJ compound jamming, and LFMJ and PPNJ compound jamming remain above $85\%$ for the same range of JNR. Additionally, the recognition accuracies for the other four types of compound jamming also maintain a commendable performance, nearly always exceeding $70\%$ from $-20$ dB to $10$ dB. These results underscore the comprehensive effectiveness of ACSNet in classifying various combinations of compound interference under diverse signal conditions.
Notably, as the JNR increases, a general upward trend in the OA for all nine types of compound interference is observed. This trend is particularly significant when the JNR surpasses $-10$ dB, at which point the OA for all types of interference consistently reaches or exceeds $92\%$. Moreover, at JNR levels of $-8$ dB and higher, the OA for all nine types nears or achieves $100\%$, indicating near-perfect classification performance. The OA for STJ and PPNJ compound jamming notably exhibits a rapid increase with rising JNR, improving from $92\%$ at $-10$ dB to $98\%$ at $-9$ dB.
These observations confirm that ACSNet not only provides high accuracy in interference recognition across a wide range of JNR settings, but also significantly improves its performance as the signal-to-interference ratio becomes more favorable. In addition, this analysis not only demonstrate the model’s adaptability to different levels of interference but also offers valuable insights for optimizing system design to ensure high efficiency and reliability in dynamically changing signal environments.

\begin{figure*}[ht]
\centering
\subfigure[]{
\begin{minipage}[t]{0.47\linewidth}
\includegraphics[width=3.5in]{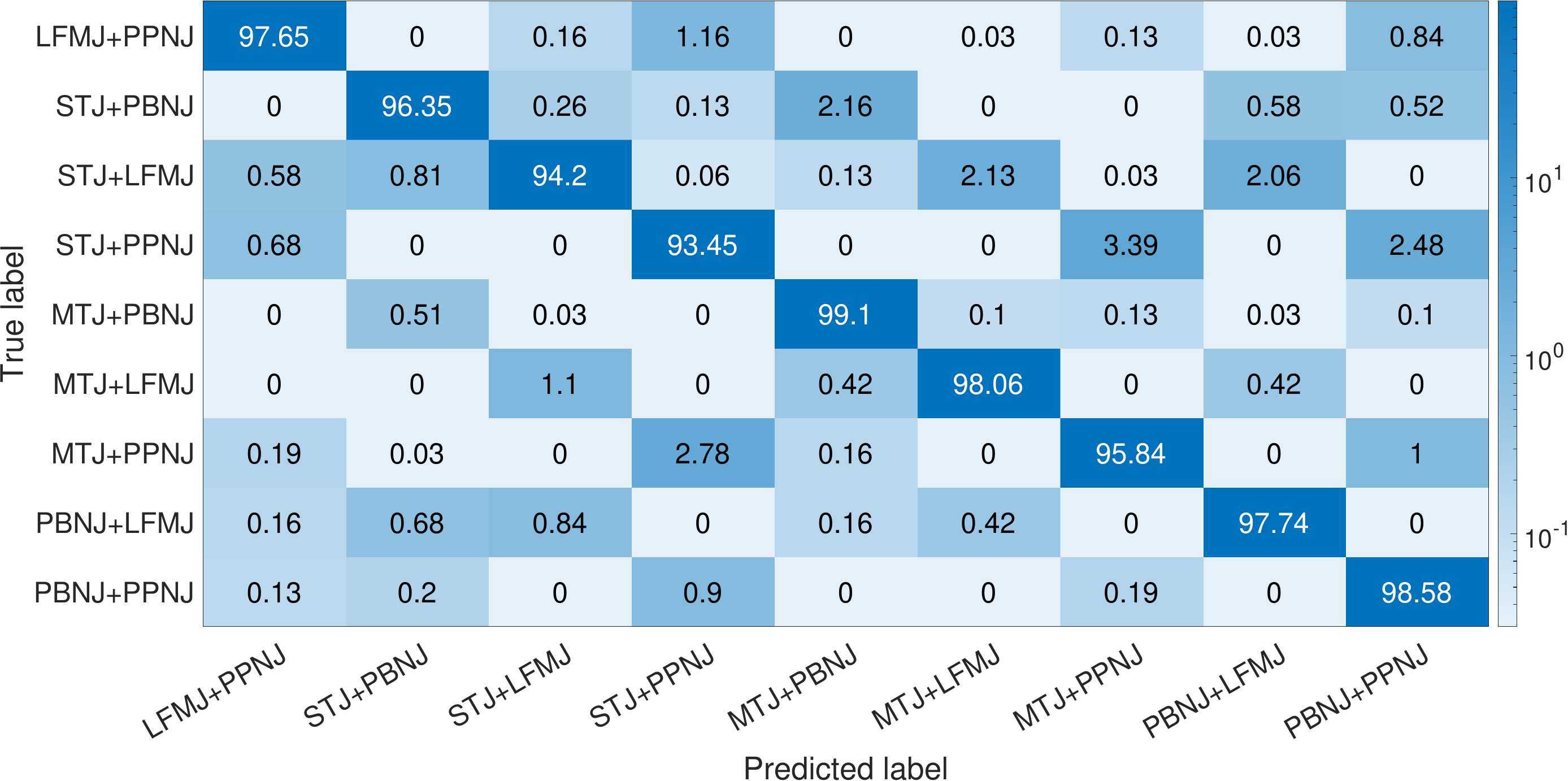}
\end{minipage}
\label{fig:5-1}
}
\subfigure[]{
\begin{minipage}[t]{0.47\linewidth}
\includegraphics[width=3.5in]{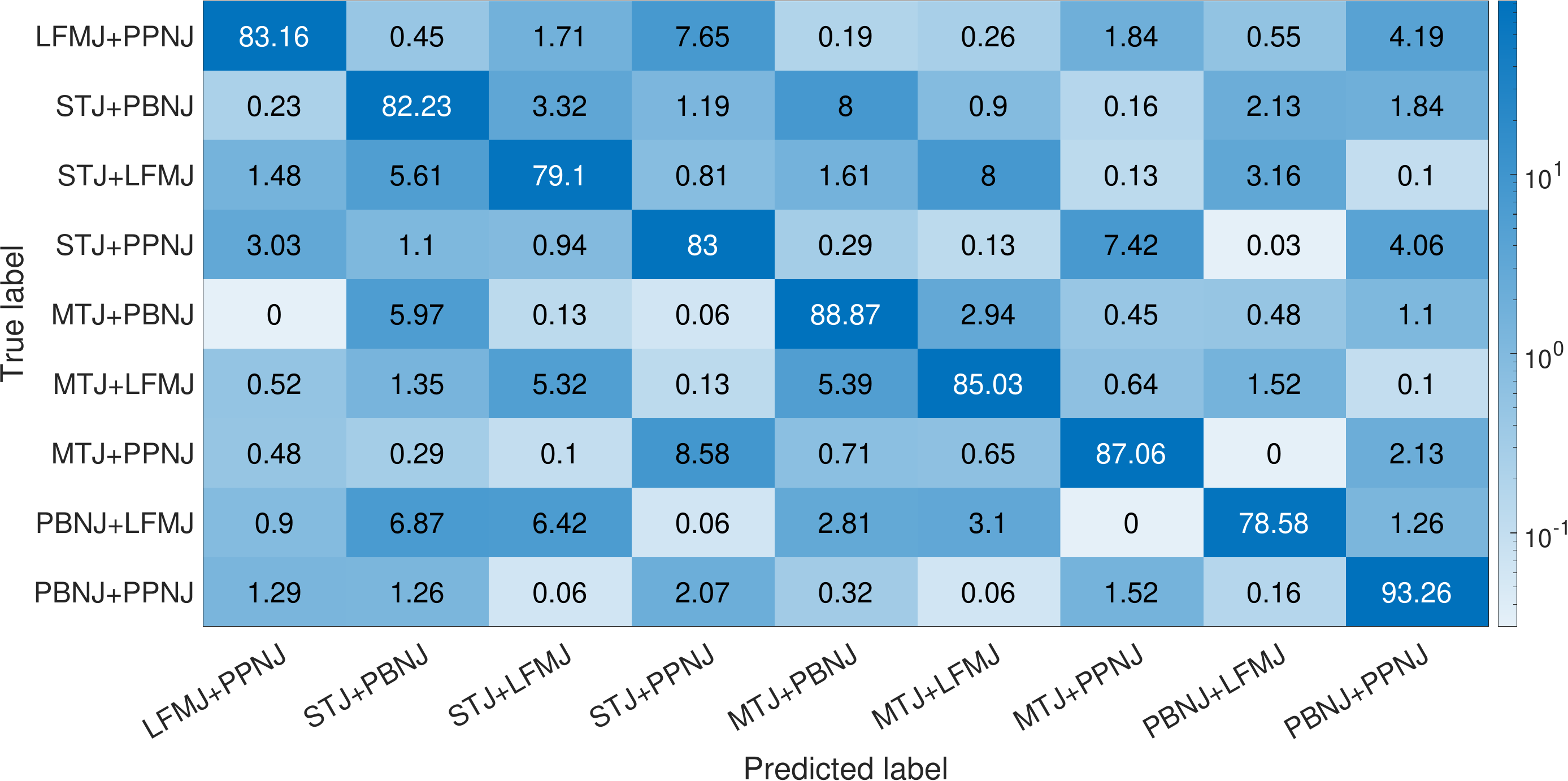}
\end{minipage}
\label{fig:5-2}
}
\subfigure[]{
\begin{minipage}[t]{0.47\linewidth}
\includegraphics[width=3.5in]{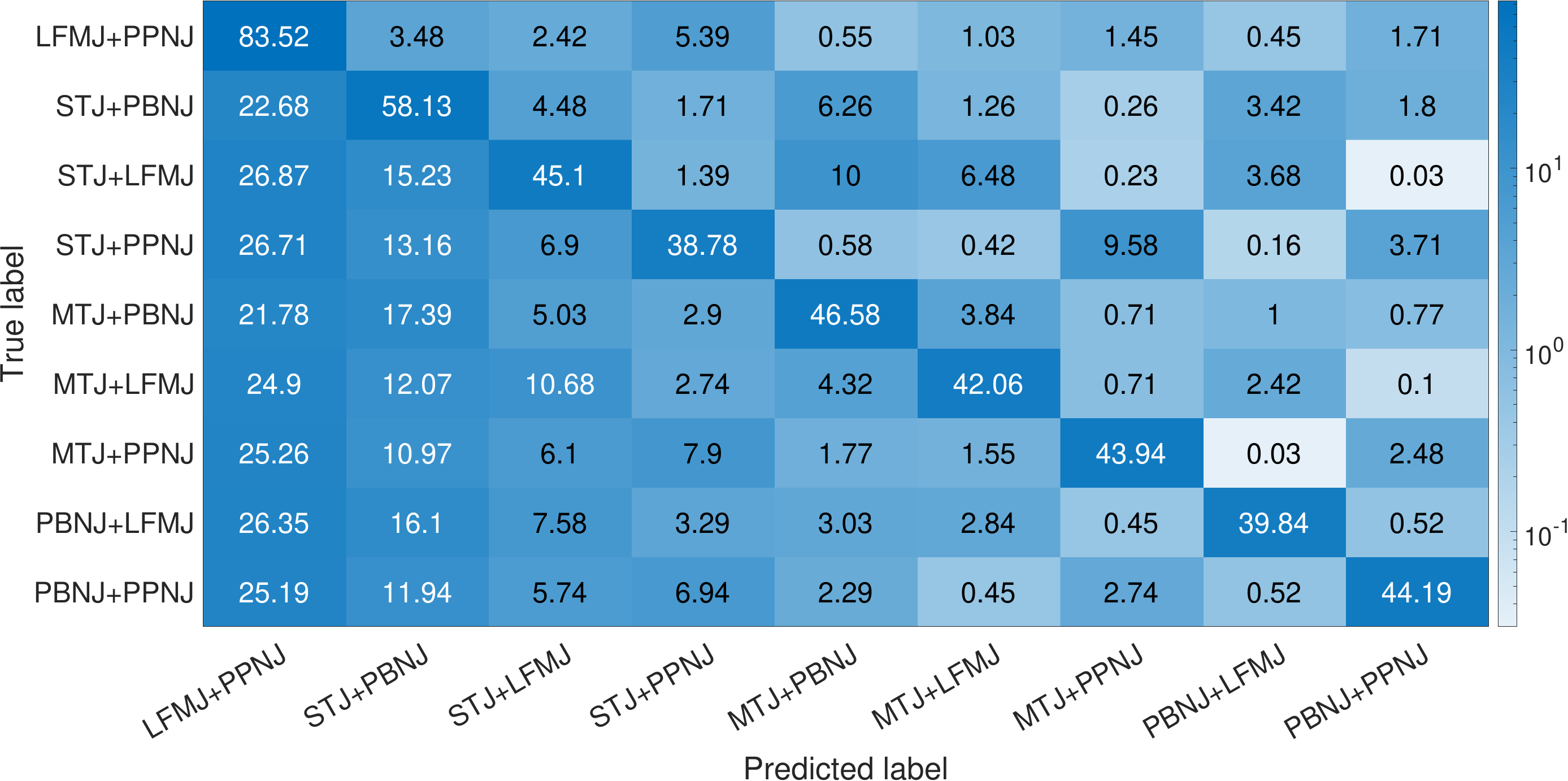}
\end{minipage}
\label{fig:5-3}
}
\subfigure[]{
\begin{minipage}[t]{0.47\linewidth}
\includegraphics[width=3.5in]{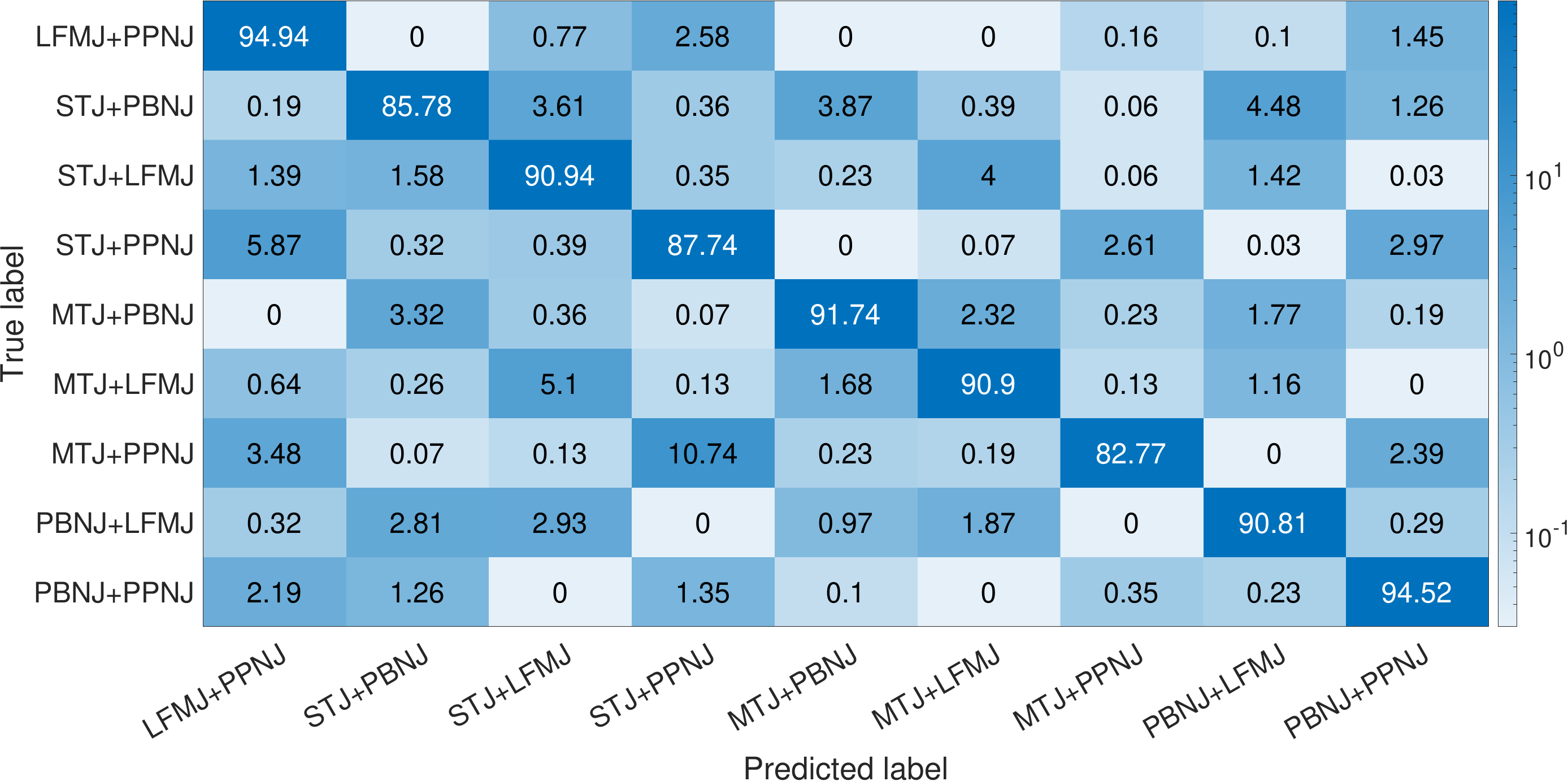}
\end{minipage}
\label{fig:5-4}
}
\caption{The recognition performance at different JNRs of different CNN models:(a) ACSNet, (b) BCNN, (c) FCNN, and (d) RCNN.}
\vspace{-5mm}
\end{figure*}

Fig. \ref{fig:5-1} delineates the confusion matrix for ACSNet, serving as a critical tool in evaluating the classification model by illustrating the relationship between predicted and true labels. This matrix offers a detailed visualization of the model’s performance across the test dataset, providing insights into both the overall and specific classification accuracies of various compound jamming types.
In the matrix, the probability of correct identification for the compound jamming combinations involving LFMJ and PPNJ, MTJ and PBNJ, MTJ and LFMJ, PBNJ and LFMJ, and PBNJ and PPNJ consistently exceeds $97\%$. These high accuracy rates underscore the effectiveness of ACSNet in correctly classifying these types of interferences. Conversely, the STJ and PPNJ compound jamming combination presents the highest challenge in terms of misclassification, with an accuracy of $93.45\%$. Notably, misclassifications include $3.39\%$ of samples being incorrectly identified as MTJ and PPNJ compound jamming, and $2.48\%$ as PBNJ and PPNJ compound jamming.
Other notable accuracies include STJ and LFMJ compound jamming at $94.2\%$, and MTJ and PPNJ compound jamming at $95.84\%$. Misclassifications among other compound interference combinations are also evident. For instance, STJ and PBNJ misclassified as MTJ and PBNJ occurs $2.16\%$ of the time, and STJ and LFMJ being misclassified as MTJ and LFMJ at a rate of $2.13\%$, with a $2.06\%$ rate of being misclassified as PBNJ and LFMJ.
Moreover, MTJ and PPNJ being misclassified as STJ and PPNJ stands at $2.78\%$. These statistics reveal that compound interferences involving STJ and MTJ are particularly susceptible to misclassification. A critical factor contributing to this issue is the potential overlap in the time-frequency representations of these signals when the number of tones in MTJ is minimal, and the frequencies of the different tones are not significantly distinct. This overlap, as illustrated in Fig. \ref{fig:1}, complicates the model’s ability to distinguish between these types of interferences effectively.
The insights derived from the confusion matrix not only highlight the model’s strengths in accurately classifying various types of compound interferences, but also pinpoint areas where the classification performance can be enhanced. In addition, understanding these dynamics is crucial for further refining the model and ensuring robust performance in diverse and complex interference environments.

\subsection{Performance Comparison with other networks}
In this section, we will compare the proposed model with three other convolutional models: BCNN\cite{morales2019jammer}, RCNN\cite{liu2019deep}, and FCNN\cite{cai2019jamming}, focusing on OA, Ka, and computational complexity, to provide a comprehensive analysis.
\begin{figure}
\centering
\includegraphics[width=0.5\textwidth]{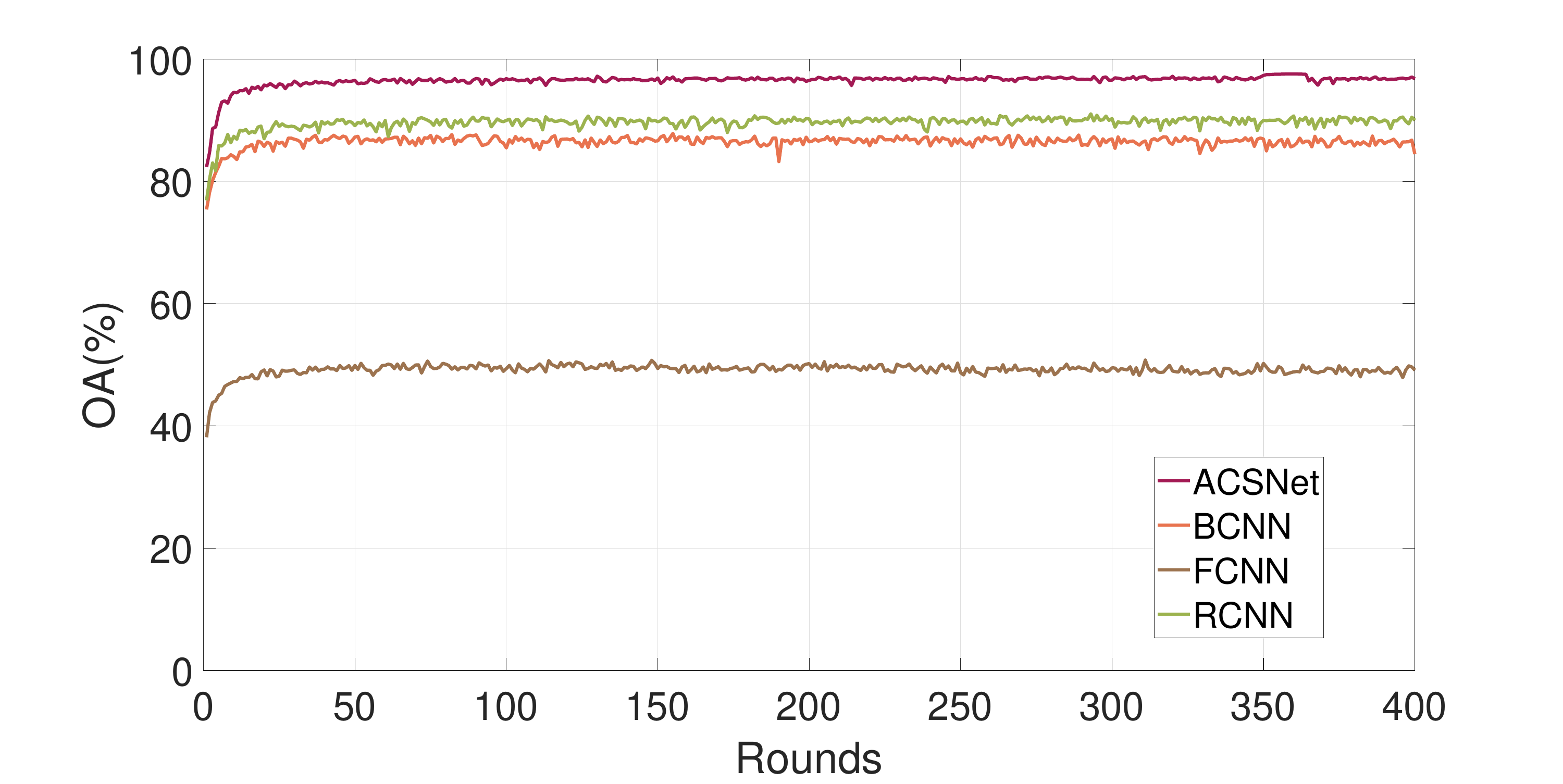}
\caption{The OAs of different CNN models in 400 epochs.}
\label{fig:6}
\end{figure}
\begin{figure}
\centering
\includegraphics[width=0.5\textwidth]{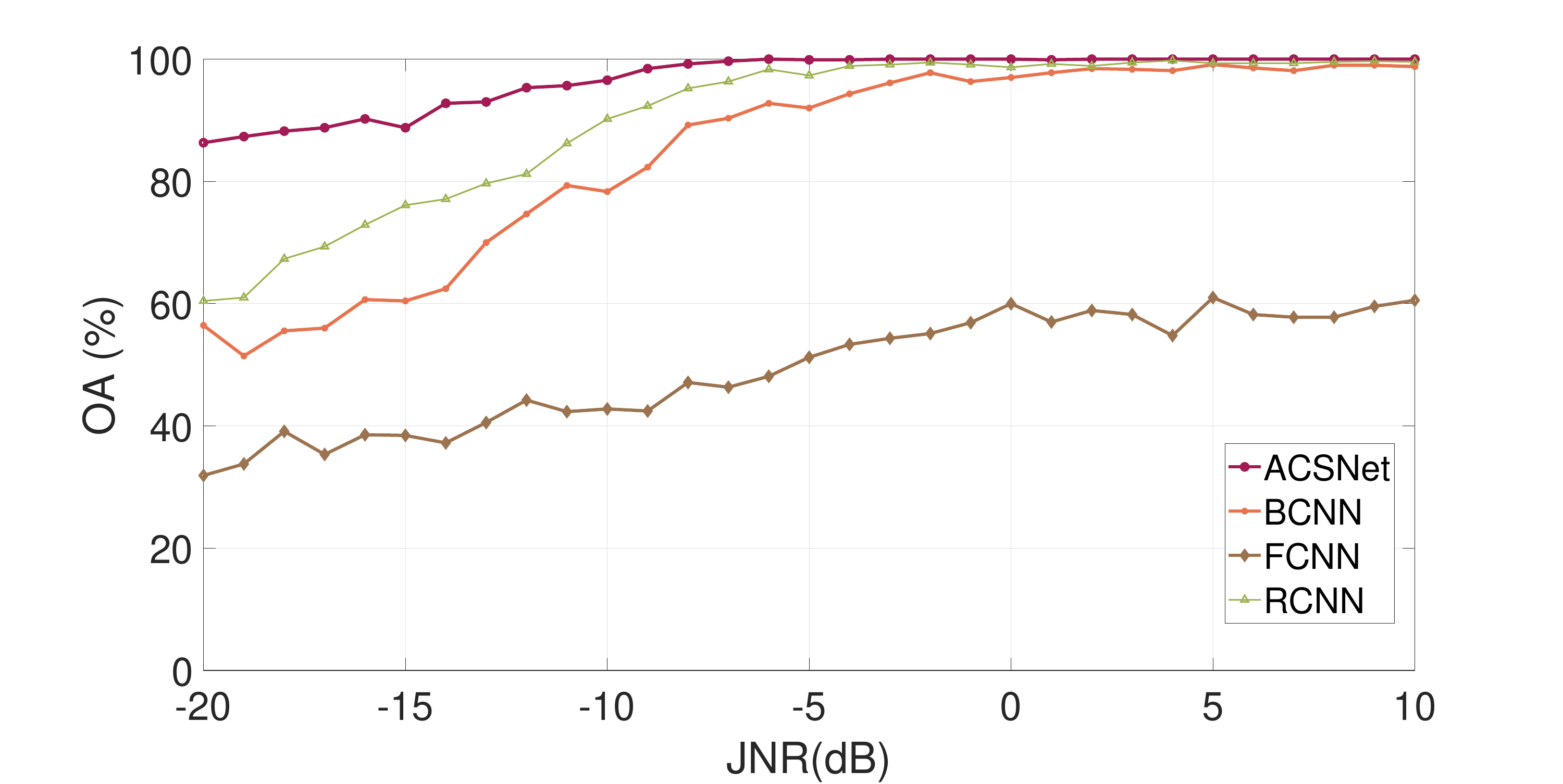}
\caption{The OAs at different JNRs of different CNN models.}
\label{fig:7}
\end{figure}
\begin{figure}
\centering
\includegraphics[width=0.5\textwidth]{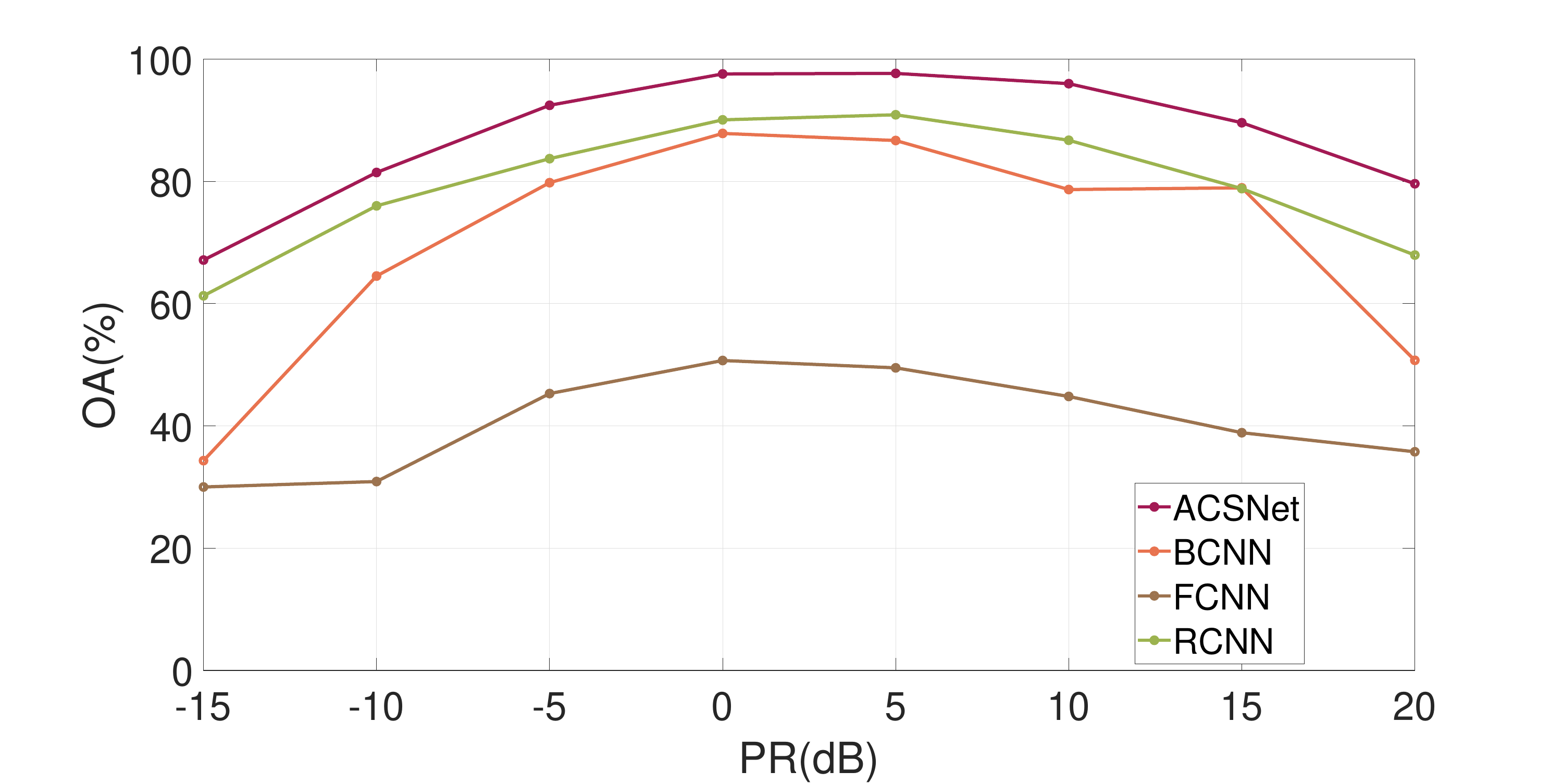}
\caption{The recognition performance at different PRs of different CNN models.}
\label{fig:8}
\end{figure}

Fig. \ref{fig:6} presents the evolution of the test set's recognition accuracy for each model across different training epochs. From the observations, it is apparent that all four models, ACSNet, RCNN, BCNN, and FCNN, converge after approximately $30$ epochs.
Among these, ACSNet demonstrates superior performance, achieving the highest recognition accuracy at $97.58\%$, which underscores its robust capability in classifying compound interference effectively. Following ACSNet, RCNN achieves a commendable accuracy of $91.06\%$, while BCNN records $87.85\%$. In contrast, FCNN shows significantly lower performance, with the lowest accuracy at only $50.70\%$. This variance in model performance highlights the effectiveness of the ACSNet architecture in achieving rapid convergence and maintaining high precision in recognizing diverse compound interference categories.
Subsequent Figs, \ref{fig:4-2}, \ref{fig:4-3}, and \ref{fig:4-4}, detail the OA at different JNRs for the BCNN, FCNN, and RCNN models, respectively.
Notably, the BCNN model exhibits a marked decline in performance under low JNR conditions.
Specifically, the recognition accuracy for STJ and LFMJ compound jamming, as well as PBNJ and LFMJ compound jamming, proves to be highly unstable.
This instability is likely attributable to the significant influence of noise, which adversely impacts the model’s ability to accurately classify these types of interference under challenging conditions.
Although RCNN and FCNN demonstrate smoother performance trajectories than that of the BCNN, their overall accuracies still lag behind that of ACSNet, particularly in scenarios characterized by low JNRs.
This comparative analysis across models provides a clear indication of ACSNet’s superior adaptability and robustness, especially in environments where noise significantly influences signal integrity.

The confusion matrices detailing the recognition performance of the other three models are presented in Fig. \ref{fig:5-2}, \ref{fig:5-3}, and \ref{fig:5-4}, respectively. These matrices provide a granular view of how each model performs across different types of compound interference, clearly indicating that their overall effectiveness is inferior to that of ACSNet.
Among these models, FCNN exhibits the weakest performance metrics, with a notably higher recognition accuracy of $83.52\%$ only for LFMJ and PPNJ compound interference. For all other types of interference, FCNN’s probabilities of correct identification fall below $60\%$, highlighting significant challenges in effectively classifying more complex or noise-sensitive interference types.
BCNN demonstrates a moderate improvement over FCNN, with the probabilities of correctly identifying various interference types generally exceeding $78\%$. This indicates a better, albeit not optimal, capacity to handle diverse interference scenarios compared to FCNN.
RCNN, among the three, shows the most promising results, with the recognition accuracy for LFMJ and PPNJ compound jamming reaching as high as $94.94\%$. This model appears to more adeptly manage the challenges presented by these interference types. However, similar to ACSNet, RCNN also exhibits vulnerability to misclassifications, particularly with compound interferences involving STJ and MTJ. This pattern suggests a common issue across models in distinguishing between these specific types of interferences, possibly due to overlapping spectral characteristics or similar temporal features within the jamming signals.
Overall, while RCNN achieves the best recognition performance among the three alternative models, all still underperform compared to ACSNet. This comparative analysis not only highlights the superior classification capabilities of ACSNet but also reinforces the model’s robustness and effectiveness in dealing with a broad spectrum of compound interferences. The lower performance metrics of the other models underscore the advanced design and optimization inherent in ACSNet, further validating its utility as a highly effective tool for interference classification in complex signal environments.

Fig. \ref{fig:7} illustrates the OA trends of different CNN models across a range of JNRs on the test dataset. The data clearly demonstrates a general improvement in recognition accuracy as JNR increases for all models. At the lower spectrum of JNR, specifically at -20 dB, ACSNet significantly outperforms its counterparts, achieving a recognition accuracy of 86.33\%. This represents a substantial improvement of 25.89\%, 29.89\%, and 54.44\% compared to RCNN, BCNN, and FCNN respectively. Such a marked difference highlights ACSNet’s robust capability in handling extremely noisy environments.

As JNR improves to $-18$ dB, ACSNet continues to maintain a lead with accuracy levels exceeding the other three models by over $20\%$. By the time the JNR reaches $-14$ dB, ACSNet’s accuracy surpasses $92\%$, while the recognition accuracies for RCNN, BCNN, and FCNN lag behind at $77.11\%$, $62.44\%$, and $37.22\%$, respectively. This trend continues up to $-10$ dB, where ACSNet and RCNN both exceed 90\% OA, but BCNN and FCNN remain below $80\%$.
A critical observation is made when the JNR level is $-9$ dB or higher; ACSNet’s OA nearly reaches $100\%$, significantly outperforming the other models, which do not exceed $93\%$. This significant difference becomes even more pronounced in extremely low noise conditions, specifically at JNRs greater than or equal to $2$ dB, where the OAs for ACSNet, RCNN, and BCNN approach $100\%$. In contrast, FCNN continues to demonstrate much lower accuracy, remaining below $62\%$.
This comprehensive performance analysis underscores the superior efficiency of ACSNet, especially at lower JNR levels. The model’s ability to capture and classify subtle features of compound interference under varied noise conditions not only testifies to its robust design, but also emphasizes its potential for practical applications where noise factors heavily. ACSNet’s consistent top-tier performance across all tested JNRs further establishes it as the optimal choice among the evaluated models, ensuring reliable and precise interference recognition.

The comparative analysis of the OA and Ka for the four network models is succinctly summarized in Table \ref{Table:2}. This table reveals that ACSNet stands out with the highest OA among the models, registering a remarkable $7.51\%$, $9.73\%$, and $46.88\%$ higher than RCNN, FCNN, and BCNN respectively. Such a significant margin not only underscores ACSNet’s superior capability in handling compound interference classification, but also highlights its efficiency in diverse testing scenarios.
In addition to its leading OA, ACSNet also demonstrates an outstanding Ka($\times 100$) of $97.29$, which further exceeds the performance of RCNN, FCNN, and BCNN by $7.4$, $10.7$, and $49.44$, respectively. The Ka, a measure of the agreement between the predicted and actual classifications adjusted for chance, offers a deeper insight into the reliability and precision of the classification results. ACSNet’s high Ka value confirms its exceptional classification accuracy and underscores its robustness and reliability in producing dependable results.
\begin{table}[h]
\caption{The OA and Ka of the four networks} \label{Table:2}
\centering
\begin{tabular}{ccccc}
\toprule  
& ACSNet &RCNN& BCNN&FCNN \\ \midrule  
OA$(\%)$&97.58&90.07&87.85&50.70\\ 
\midrule
Ka$(\times100)$&97.29&89.89&86.59&47.85\\
\bottomrule  
\end{tabular}
\end{table}

To assess the computational efficiency and resource utilization of the network models, we meticulously analyze several critical aspects: floating-point operations (FLOPs), the number of learnable parameters, and the inference time.
As detailed in \cite{qu2020jrnet}, the network architectures under consideration primarily comprise convolutional layers and linear layers, which are the main contributors to computational load.
The methodologies for calculating the FLOPs for these layers are essential for a standardized assessment of computational complexity and are outlined as follows:
\begin{equation}\label{eq:26}
    FLOP_{s_c} = \sum_{l=1}^{D} 2{M_l}^2{k_l}^2C_{l-1}C_l,
\end{equation}
\begin{equation}\label{eq:27}
    FLOP_{s_l} = \sum_{l=1}^{D} (2C_{l-1}-1)C_l,
\end{equation}
where $FLOP_{s_c}$ and $FLOP_{s_l}$ represent the FLOPs of the convolutional layer and the linear layer, respectively. $C$ and $D$ denote the numbers of layers and channels, respectively. $M$ and $k$ are the lengths of feature maps and kernels of $l$-th layer, respectively.
\begin{table}[h]
\caption{The computational complexity of the four networks} \label{Table:3}
\centering
\begin{tabular}{cccc}
\toprule  
Networks& Parameters(M) &FLOPs(G)& Time(ms) \\ \midrule  
ACSNet&1.65&0.18&1.713\\ 
\midrule
RCNN&1.32&0.35&0.792\\ 
\midrule
BCNN&0.49&0.10&0.527\\ 
\midrule
FCNN&8.39&0.05&0.568\\ 
\bottomrule  
\end{tabular}
\end{table}
As detailed in Table \ref{Table:3}, the computational complexity metrics for the four network models are presented alongside inference time measurements, based on the specific hardware platform utilized in the simulations.
Notably, ACSNet, with a total of $1.65M$ learnable parameters, possesses more parameters than RCNN ($0.33M$ fewer) and BCNN ($1.16M$ fewer).
Despite having a larger parameter count than these models, ACSNet still maintains a significantly lower count compared to FCNN, which has a substantial $8.39M$ parameters.
Furthermore, the FLOPs for ACSNet amount to $0.18G$, which is markedly lower than the $0.35G$ recorded for RCNN, indicating a more efficient computational structure. Although ACSNet’s inference time is marginally longer by approximately $1$ millisecond compared to the other networks, this slight increase is outweighed by its superior recognition performance.

The simulations conducted previously maintained a PR of $0$ dB, ensuring equal power distribution among the components of compound jamming signals. This uniform power setting serves as a baseline for understanding model performance under standardized conditions. However, to address the complexities of real-world scenarios where jamming signals often vary in intensity, further simulations were executed with PR values ranging from $-15$ dB to $25$ dB, as depicted in Fig. \ref{fig:8}.

The results from these simulations reveal a general trend where the recognition rates for all models decrease as the PR deviates from $0$ dB, whether increasing or decreasing. Interestingly, the decline in recognition accuracy is more pronounced when PR exceeds $0$ dB compared to when it is below $0$ dB.
Notably, both ACSNet and RCNN exhibit a slight increase in recognition accuracy at a PR of $5$ dB, higher by about $0.09\%$ and $0.83\%$, respectively, compared to their performance at PR $=0$ dB.
This improvement might be attributed to the enhanced distinguishability of the features from the higher-power jamming signal, which aids in more effective recognition. Such a scenario underscores the models’ sensitivity to variations in signal power, where a slight increase in disparity can enhance feature detectability.
Among the models, BCNN demonstrates the most significant fluctuations in accuracy with changes in PR, indicating a higher sensitivity to power distribution variations within the jamming signals. Conversely, FCNN shows a more gradual change in accuracy across different PR values but consistently records lower overall accuracy. ACSNet and FCNN exhibit similar rates of variation in performance; however, ACSNet consistently achieves superior recognition accuracy. Remarkably, ACSNet maintains an accuracy above $90\%$ for PR values ranging from $-5$ dB to $15$ dB, showcasing its robust capability to extract relevant features even in the presence of dominant high-power jamming components.
\section{Conclusion}
In this paper, we have presented ACSNet, a CNN-based model specifically developed to recognize nine prevalent types of additive compound jamming in GNSS systems. The model’s efficacy has been demonstrated through extensive simulations, which highlighted its capability to maintain high recognition accuracy across varying JNR conditions.
Notably, ACSNet has achieved an accuracy of $86.33\%$ at a JNR of $-20$ dB and approached $100\%$ accuracy for JNR values greater than or equal to $-9$ dB.
Furthermore, the OA and Ka($\times100$) of ACSNet, calculated as $97.58\%$ and $97.29$, respectively, have significantly surpassed those of the competing convolutional networks, underscoring its superior recognition performance and precision.
The promising results from these simulations have demonstrated that ACSNet not only excels in recognizing additive compound jamming scenarios, but also has potential applicability to both additive and multiplicative compound jamming contexts. Building on this, future research will aim to further explore and validate ACSNet’s effectiveness across these broader applications, enhancing its utility for practical deployment in complex signal environments.

\bibliographystyle{ieeetr}
\bibliography{reference}

\end{document}